\documentclass[journal]{IEEEtran}
\usepackage{cite}
\usepackage[cmex10]{amsmath}
\usepackage{algorithm}
\usepackage{graphicx}
\usepackage{color}
\usepackage{mdwmath}
\usepackage{mdwtab}
\usepackage{amssymb}

\DeclareMathOperator*{\E}{\mathbb{E}}

\begin{document}

\title{Spatial Sensing and Cognitive Radio Communication in the Presence of A $K$-User Interference Primary Network}

\author{Ardalan~Alizadeh,~\IEEEmembership{Student~Member,~IEEE,}   ~Hamid~Reza~Bahrami,~\IEEEmembership{Member,~IEEE,}
~Mehdi~Maleki,~\IEEEmembership{Student~Member,~IEEE,} and~Shivakumar Sastry,~\IEEEmembership{Senior~Member,~IEEE}% 
\thanks{Manuscript received January 5, 2014; revised May 8, 2014 and July 18, 2014.}%
\thanks{The authors are with the Department of Electrical and Computer Engineering, The University of Akron, Akron, OH, 44325 USA e-mail: \{aa148, hrb, mm158, ssastry\}@uakron.edu.}
\thanks{ \copyright 2014 IEEE. Personal use of this material is permitted. Permission from IEEE must be obtained for all other uses, in any current or future media, including reprinting/republishing this material for advertising or promotional purposes, creating new collective works, for resale or redistribution to servers or lists, or reuse of any copyrighted component of this work in other works.}}

\markboth{IEEE JOURNAL ON SELECTED AREAS IN COMMUNICATIONS,~Vol.~X, No.~X, April~2014}%
{Alizadeh \MakeLowercase{\textit{et al.}}: Spatial Sensing and Cognitive Radio Communication in the Presence of A $K$-User Interference Primary Network}

\maketitle

\begin{abstract}
We study the feasibility of cognitive radio (CR) communication in the presence of a $K$-user multi-input multi-output (MIMO) interference channel as the primary network. Assuming that the primary interference network has unused spatial degrees of freedom (DoFs), we first investigate the sufficient condition on the number of antennas at the secondary transmitter under which the secondary system can communicate while causing no interference to the primary receivers. We show that, to maximize the benefit, the secondary transmitter should have at least the same number of antennas as the spatial DoFs of the primary system. We then derive the secondary precoding and decoding matrices to have zero interference leakage into the primary network while the signal-to-interference plus noise ratio (SINR) at the secondary receiver is maximized. As the success of the secondary communication depends on the availability of unused DoFs, we then propose a fast sensing method based on the eigenvalue analysis of the received signal covariance matrix to determine the availability of unused DoFs or equivalently spatial holes. Since the proposed fast sensing method cannot identify the indices of inactive primary streams, we also provide a fine sensing method based on the generalized likelihood ratio test (GLRT) to decide the absence of individual primary streams. Simulation results show that the proposed CR sensing and transmission scheme can, in practice, provide a significant throughput while causing no interference to the primary receivers, and that the sensing detects the spatial holes of the primary network with high detection probability.
\end{abstract}


\IEEEpeerreviewmaketitle

%%%%%%%%%%%%%%%%%%%%%%%%%%%%%%%%%%%%%%%%%%%%%%%%%%%%%%%%%%%%%
\section{Introduction}
\IEEEPARstart{T}{he} current explosion of information and demand for high speed data communication call for novel solutions to utilize the radio resources more efficiently. The cognitive radio (CR) paradigm aims to mitigate this spectrum crunch by opportunistically exploiting unused licensed spectrum bands that are allocated to the primary communications systems \cite{Haykin2005}. Since it is a key design objective to not compromise the performance of the primary system, CR schemes are designed to improve spectrum utilization at little or no extra cost to the licensed users. The design of efficient CR schemes also depends on the structure and the architecture of the primary system. For examples, different spectrum sensing and CR communication schemes are necessary when the primary system is a point-to-point network \cite{Axell}, or a cooperative relay network with single \cite{Alizadeh2010,Budhathoki2012,Alizadeh2011} or multiple antennas \cite{K.R.Budhathoki2013}.

Any efficient opportunistic CR network must accurately detect the presence or absence of communication in the primary system. Several sensing schemes including energy detection, matched filtering and cyclostationary detection (see e.g. \cite{Zeng2010} and the references therein) have been proposed in the literature for detecting primary spectral holes. Although the energy detection method does not require \textit{a priori} information of the primary signal \cite{Alizadeh2011IEICE}, it is not optimal for detecting correlated signals. To overcome the shortcomings of energy detection, eigenvalue based sensing schemes have been suggested \cite{Zeng2008,Zeng2009a}. Further, especially for the case of multi-antenna CR nodes, the generalized likelihood ratio test (GLRT) has been proposed to utilize the eigenvalues of the sample covariance matrix of the received signal vector without having \textit{a priori} knowledge of the primary users' signals \cite{Zhang2010,Lim2008}. In this method, the primary users' signals to be detected occupy a subspace of dimension strictly smaller than that of the observation space \cite{Zhang2010}. 

In this paper, we propose opportunistic spectrum usage of a primary $K$-user interference network operating under the \emph{interference alignment} (IA) scheme. Unlike conventional CR schemes that search for spectral holes, our proposed CR sensing and communication scheme identifies and utilizes \textit{spatial holes} of the primary interference network. We consider a simple system model composed of a point-to-point multi-input multi-output (MIMO) secondary system, while the primary network consists of $K$ pairs of multi-antenna nodes. We assume that the primary is able to use the whole achievable \emph{degrees of freedom} (DoFs) with the recently introduced idea of the $K$-user interference alignment  \cite{Cadambe2008,Jafar2010}, in which each transmitter sends at least one stream of data. Our aim is to design a CR scheme that utilizes the unused DoFs of the primary system for secondary transmission at no extra cost to the primary. This is done by designing the secondary precoder and decoder matrices such that  no interference is imposed to the primary network, while the signal-to-interference plus noise ratio (SINR) at the secondary receiver is maximized. As the success of this CR scheme depends on successful detection of the spatial holes in the primary system, we derive a spatial sensing method by taking advantage of eigenvalue based and GLRT sensing schemes in two stages to firstly detect the presence of spatial hole(s) and then to find the index of inactive primary stream(s). In practice, the proposed CR system first detects the presence of spatial holes by performing a low complexity eignenvalue based sensing. If it detects the presence of spatial holes, the CR then proceeds with finding the indices of the unused DoFs based on the primary channel state information (CSI), which are then used to design the secondary precoder and decoder matrices. The secondary precoder and decoder matrices are such that they maximize the secondary SINR while causing no interference to the primary receivers. Even though the primary and secondary transmissions are designed for the considered system model, the proposed sensing scheme is general and applicable to any scenario that involves spatial sensing.

\subsection{Related Works}
The problem of space pooling has been partly studied in \cite{Perlaza2008} and \cite{Perlaza2010} for point-to-point MIMO primary and secondary systems. In these papers, authors present an opportunistic scheme to utilize the unused eigenmodes of the primary channel, and introduce pre- and post-processing schemes. Recently, using the IA techniques in CR networks has been considered {\cite{Amir2011,Zhou2010,Chen2012,Koo2012,Abdelhamid2012,Xu2013,Wu2010,Du2011,Abdelhamid2013,Guler2013}}. In \cite{Amir2011}, authors consider a point-to-point MIMO  primary network and propose an iterative algorithm to efficiently design the precoding and decoding matrices for IA in the secondary network. Based on a similar network setup, \cite{Zhou2010} studies the impact of propagation delay on the DoF of the CR system. In \cite{Chen2012}, an outer bound for the total DoF is derived. The achievable DoF when the primary network performs interference suppression is discussed in \cite{Koo2012}. With the same network model, authors in \cite{Abdelhamid2012} provide pre- and post-processing designs of the CR IA network to maintain a target rate for the primary network while maximizing the rate of the secondary link. Recently in \cite{Xu2013}, the problem of cooperative spectrum leasing based on a game theoretical approach in an IA network composed of primary and secondary users has been studied. In \cite{Wu2010,Du2011,Abdelhamid2013}, authors study the use of IA in underlay CR systems, while the application of IA in CR femtocell networks has been studied in \cite{Guler2013}.
\subsection{Contributions}
The main contributions of this paper are:
\begin{itemize}
\item the condition on the number of antennas at the secondary link to exploit the spatial holes in the primary system without causing interference to the primary system as well as the structures of the secondary precoding and decoding matrices.
\item a fast method for detecting spatial holes in the primary system; this method is based on the eigenvalues of the received covariance matrix, and is a coarse sensing scheme that detects the presence of unused or inactive DoFs in the primary system.
\item and finally, a fine sensing method that identifies the inactive primary streams; this method is applied after the coarse sensing to accurately determine the index set of inactive data streams in the primary system.
\end{itemize}
The remainder of this paper is organized as follows. Section II introduces the system model and problem formulation. Section III presents the method for alignment of the secondary transmission in the CR system for opportunistic usage of the available spatial holes in the IA primary system. In Section IV and Section V, we propose the two sensing techniques for coarse and fine detection of the null spaces, respectively. Numerical results are provided in Section VI to verify the effectiveness of the proposed scheme. Finally, we conclude the paper in Section VII.

Throughout the paper, we use lower case letters for scalars. Vectors and matrices are generally shown with uppercase italic and bold letters, respectively. $\textbf{I}_d$ denotes an identity matrix of size $d$. $\textbf{0}_{M\times N}$ indicates an $M \times N$ all zero matrix.  $\E\{\textbf{A}\}$ and $\text{Tr}\{\textbf{A}\}$ are expectation and trace of matrix $\textbf{A}$. $\textbf{A}^{T}$ and $\textbf{A}^{\dag}$ are the transpose and conjugate transpose of matrix $\textbf{A}$, respectively. $\textbf{A}_{\parallel d}$ represents the $d^{th}$ column of matrix $\textbf{A}$. $\text{E}_{d}\left[\textbf{A}\right]$ stands for the eigenvector corresponding to the $d^{th}$ smallest eigenvalue of $\textbf{A}$.
%%%%%%%%%%%%%%%%%%%%%%%%%%%%%%%%%%%%%%%%%%%%%%%%%%%%%%%%%%%%%
\section{System Model}
%%%%%%%%%%%%%%%%%%%%%%%%%%%%%%%%%%%%%%%%%%%%%%%%%%%%%%%%%%%%%
\subsection{Overview of the Scenario}
\begin{figure}[!t]
\centering
\includegraphics[width=8.7cm]{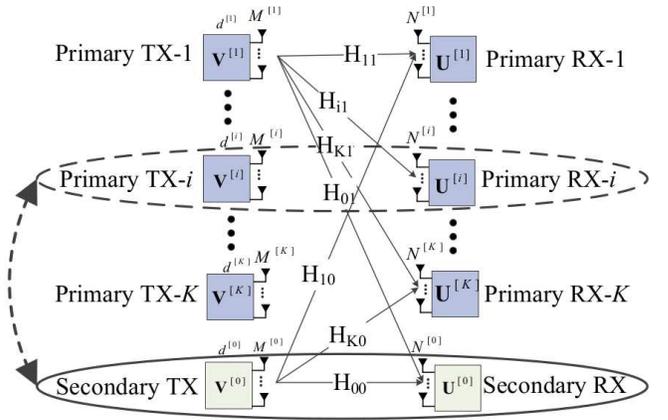}
\caption{An opportunistic cognitive radio system in the presence of a $K$-user interference alignment primary network. When the $i^{th}$ primary stream is inactive, the secondary system is able to utilize the unused DoFs for its own transmission.}
\vspace{-0.5cm}
\label{oppor}
\end{figure}

We consider a $K$-user MIMO interference channel shown in Fig.~\ref{oppor} as the primary system, where the IA is done at the primary transmitters having full CSI. In this network, each transmitter transmits one or a few streams of data to its corresponding receiver. However, we assume that during some transmission intervals one or some of the primary transmitters does not have any data to transmit, and therefore, some DoFs of the primary system remain unused at those time intervals. We consider a pair of multi-antenna secondary nodes to opportunistically utilize the unused DoFs without causing interference to the primary system. 

%%%%%%%%%%%%%%%%%%%%%%%%%%%%%%%%%%%%%%%%%%%%%%%%%%%%%%%%%
%%%%%%%%%%%%%%%%%%%%%%%%%%%%%%%%%%%%%%%%%%%%%%%%%%%%%%%%%

It is assumed that the secondary system has full CSI of all the primary links, primary to secondary, secondary to primary, as well as the link between the secondary transmitter and receiver. Although acquiring full CSI at the secondary system is difficult in practice, there are a number of ways to make this possible.  Based on the channel reciprocity, CR nodes are able to estimate the channel information from the primary to secondary nodes and vice versa using pilot signals of the primary network. A critical assumption in the realization of a distributed IA network is the availability of robust feedback channels \cite{Thukral2009}. In addition, in the proposed scheme, the CR network requires the knowledge of precoder and decoder matrices of the primary network. A practical way to obtain this information in the CR network using cognitive pilot channel (CPC) has been introduced \cite{Perez2007}, and currently standard organizations are considering using such channels to broadcast useful information of primary to available secondary networks for opportunistic or underlay utilization of the spectrum \cite{ETSI2009}.
\subsection{Primary IA network}
We assume that in the primary network, the $k^{th}$ transmitter and receiver pair have $M^{[k]}$ and $N^{[k]}$ antennas, respectively. The DoF for the signal of the $k^{th}$ pair is defined by $d^{[k]}\leq \min(M^{[k]},N^{[k]})$, $k\in \mathcal{K}$ where $\mathcal{K} \triangleq \{1,2,...,K\}$ is the set of all primary pairs \cite{Cadambe2008,Jafar2010}. The receive signal at the $k^{th}$ primary receiver can be written as:
\begin{equation}
Y^{[k]}=\sum_{l=1}^{K}\textbf{H}^{[kl]}X^{[l]}+Z^{[k]}, ~~~\forall k \in \mathcal{K}
\end{equation}
where $X^{[l]}$ is the $M^{[l]}\times 1$ transmit signal vector of the $l^{th}$ primary transmitter, $Y^{[k]}$ is the $N^{[k]}\times 1$ receive vector, and $Z^{[k]}$ is the complex additive white Gaussian noise (AWGN) vector at the $k^{th}$ receiver with independent and identically distributed (i.i.d) circularly symmetric complex Gaussian entries with zero mean and variance $\sigma_z^2$. $\textbf{H}^{[kl]}$ is the $N^{[k]} \times M^{[l]}$ matrix of channel coefficients between the $l^{th}$ transmitter and the $k^{th}$ receiver as shown in Fig. \ref{oppor}. For simplicity, it is assumed that all channel matrices are full rank; however, the proposed scheme can be easily extended to the case of rank deficient matrices. 
\subsubsection{Precoder at primary}
The signal vector transmitted by the $k^{th}$ transmitter can be written as:
\begin{equation}
X^{[k]}=\sum_{d=1}^{d^{[k]}}\textbf{V}^{[k]}_{\parallel d}\bar{X}^{[k]}_d=\textbf{V}^{[k]}\bar{X}^{[k]},
\end{equation}
where $\bar{X}^{[k]}$  is a $d^{[k]} \times 1$ vector, $\bar{X}^{[k]}_d$ is the $d^{th}$ entry of vector $\bar{X}^{[k]}$ and $\textbf{V}^{[k]}$ is an $M^{[k]} \times d^{[k]}$ precoder matrix whose columns are the orthogonal basis for the transmitted signal space of the $k^{th}$ transmitter \cite{Jafar2010}. We assume that the average transmit power at the $k^{th}$ transmitter is $\E[||X^{[k]}||^2]=p^{[k]}$ and that the power is allocated uniformly to all data streams. 
\subsubsection{Decoder at primary}
The interference suppression at the primary receiver can be utilized to construct an $N^{[k]} \times d^{[k]}$ decoder matrix $\textbf{U}^{[k]}$ with orthogonal columns to minimize the interference in the desired signal subspace at the $k^{th}$ receiver. Therefore, the receive signal after decoder for the $k^{th}$ receiver is:
\begin{equation}
\bar{Y}^{[k]}=\textbf{U}^{[k]\dag}Y^{[k]}.
\end{equation}
\subsubsection{Feasibility of alignment}
If the interference is aligned into the null space of $\textbf{U}^{[k]}$, then the following condition must be satisfied:
\begin{equation}
\textbf{U}^{[k]\dag}\textbf{H}^{[kj]}\textbf{V}^{[j]}=\textbf{0}_{d^{[k]} \times d^{[j]}}, \forall j \neq k,
\end{equation}
\begin{equation}
\text{rank}(\textbf{U}^{[k]\dag}\textbf{H}^{[kk]}\textbf{V}^{[k]})=d^{[k]}, \forall k \in \mathcal{K},
\end{equation}
while columns of the precoding and decoding matrices construct orthogonal basis sets; i.e.:
\begin{equation}
\textbf{V}^{[k]}:M^{[k]} \times d^{[k]}, \textbf{V}^{[k] \dag}\textbf{V}^{[k]}=\textbf{I}_{d^{[k]}},
\end{equation}
\begin{equation}
\textbf{U}^{[k]}:N^{[k]} \times d^{[k]}, \textbf{U}^{[k] \dag}\textbf{U}^{[k]}=\textbf{I}_{d^{[k]}}.
\end{equation}
\subsubsection{Numerical solutions}
Although closed-form solutions have been found for the IA problem in three-user interference channels \cite{Jafar2010}, the closed-form solution for the general case of $K$-user interference channel is unknown and such a problem is NP-hard. There are, however, some numerical iterative algorithms suggested in the literature to design the precoder and decoder in the case of $K$-user IA with full CSI \cite{Gomadam2011,Peters2011,Razaviyayn2012}. In this paper, and in the numerical results section, we consider the method provided in \cite{Gomadam2011} to derive the precoding and decoding matrices in the primary network. This method considers the reciprocity of the channels and assigns the eigenvector corresponding to the the smallest eigenvalue of the receive matrix as precoder. In the next step, the decoder at the receiver can be considered as the precoder of the reciprocal channel. This iteration continues until the solution converges to an IA solution.

\subsection{Secondary MIMO Link}
We consider a pair of multi-antenna transmitter and receiver as the secondary network (the pair with index $k=0$ shown in Fig. \ref{oppor}). It is assumed that the secondary network works at the same frequency and time as that of the primary network. The $M^{[0]} \times d^{[0]}$ matrix $\textbf{V}^{[0]}$ and the $N^{[0]} \times d^{[0]}$ matrix $\textbf{U}^{[0]}$ are the precoding and decoding matrices of the secondary network, respectively. The number of streams that can be transmitted by the CR system is $d^{[0]}$. Our aim is to find these two matrices to minimize the interference imposed to the primary network while maximizing the transmission rate of the secondary system.
\subsubsection{Interference leakage}
We assume that the interferences in the primary network have been aligned and that the $k^{th}$ primary transmitter has $d^{[k]}$ degrees of freedom. The interference leakage at the $k^{th}$ primary receiver due to the secondary transmission is the summation of all interferences from secondary streams to the $k^{th}$ primary receiver; i.e.:
\begin{equation}
I^{[k]}_{\text{CR}}=\text{Tr}\left[\dfrac{p^{[0]}}{d^{[0]}}\textbf{U}^{[k]\dag}\textbf{H}^{[k0]}\textbf{V}^{[0]}\textbf{V}^{[0] \dag}\textbf{H}^{[k0]\dag}\textbf{U}^{[k]}\right],
\end{equation}
where it is assumed that the total secondary transmit power, $p^{[0]}$, is allocated uniformly to all data streams.
\subsubsection{SINR at CR receiver}
The SINR of the $l^{th}$ secondary data stream at the secondary receiver can be written as \cite{Gomadam2011}:
\begin{equation}\label{SINR}
\gamma^{\rm CR}_{l}=\dfrac{p^{[0]}}{d^{[0]}}\dfrac{\textbf{U}^{[0]\dag}_{\parallel l}\textbf{H}^{[00]}\textbf{V}^{[0]}_{\parallel l}\textbf{V}^{[0]\dag}_{\parallel l}\textbf{H}^{[00]\dag}\textbf{U}^{[0]}_{\parallel l}}{\textbf{U}^{[0]\dag}_{\parallel l}\textbf{B}_{l}\textbf{U}^{[0]}_{\parallel l}},
\end{equation}
where 
\begin{align}\label{eq10}
\textbf{B}_{l}=&\sum_{j=0}^{K}\dfrac{p^{[j]}}{d^{[j]}}\sum_{d=1}^{d^{[j]}}\textbf{H}^{[0j]}\textbf{V}^{[j]}_{\parallel d}\textbf{V}^{[j]\dag}_{\parallel d}\textbf{H}^{[0j]\dag} \\ \notag
&-\dfrac{p^{[0]}}{d^{[0]}}\textbf{H}^{[00]}\textbf{V}^{[0]}_{\parallel l}\textbf{V}^{[0]\dag}_{\parallel l}\textbf{H}^{[00]\dag}+\sigma_z^2\textbf{I}_{N^{[0]}},
\end{align}
where we have not considered the possibility of successive interference cancellation in the SINR equation.
%%%%%%%%%%%%%%%%%%%%%%%%%%%%%%%%%%%%%%%%%%%%%%%%%%%%%%%%%%%%%
\section{Secondary Interference Alignment}
%%%%%%%%%%%%%%%%%%%%%%%%%%%%%%%%%%%%%%%%%%%%%%%%%%%%%%%%%%%%%
In this section, we consider the design of the secondary precoder and decoder matrices in the presence of the primary $K$-user interference network. We, first, derive the sufficient condition on the number of antennas at the secondary transmitter in order to utilize the spatial holes for its own transmission without causing interference to primary. Then, we provide the pre- and post-processing (precoding and decoding) matrices to maximize the SINR at the CR receiver while the interference leakage to the primary is zero. 
\subsection{Precoder at Secondary Transmitter}
We consider the summation of interference terms from the secondary transmitter to the primary receivers. Our aim is to design the precoder at the secondary transmitter such that the interference leakage to the primary receivers is minimized or optimally nullified. Therefore, the optimization problem to find the secondary precoder can be written as:
\begin{align} \label{precoder}
&\min_{\textbf{V}^{[0]}:M^{[0]}\times d^{[0]}} \sum_{k=1}^{K}I^{[k]}_{\text{CR}} \\ \notag
&\text{s.t.~} \textbf{V}^{[0]\dag}\textbf{V}^{[0]}=\textbf{I}_{d^{[0]}},
\end{align}
where the precoder matrix $\textbf{V}^{[0]}$ at the CR transmitter is assumed to be a unitary matrix to preserve the secondary transmit power constraint. While such a choice will minimize the interference, it does not always guarantee zero interference at each primary receiver. Instead, we shall first state the sufficient condition on the number of transmit antennas at the secondary to completely nullify the interference leakage to the primary receivers, and then propose an alternative precoder design to achieve zero interference.

\newtheorem{lemma}{Lemma}
\begin{lemma}[Sufficient condition for zero interference] The number of secondary transmit antennas is limited by 
\begin{equation}\label{eq:lemma1}
M^{[0]} \geq d^{[0]}+\sum_{i =1}^{K}d_\mathcal{A}^{[i]},
\end{equation}
where $d_\mathcal{A}^{[i]}$ is the number of active streams of the $i^{th}$ primary transmitter. 
\end{lemma}
\begin{IEEEproof}
%%%%%%%%%%%%%%%%%%%%%%%%%%%%%%%%%%%%%%%%%%%%%%%%%%%%%%%%%
\begin{figure}[!t]
\centering
\includegraphics[width=9cm]{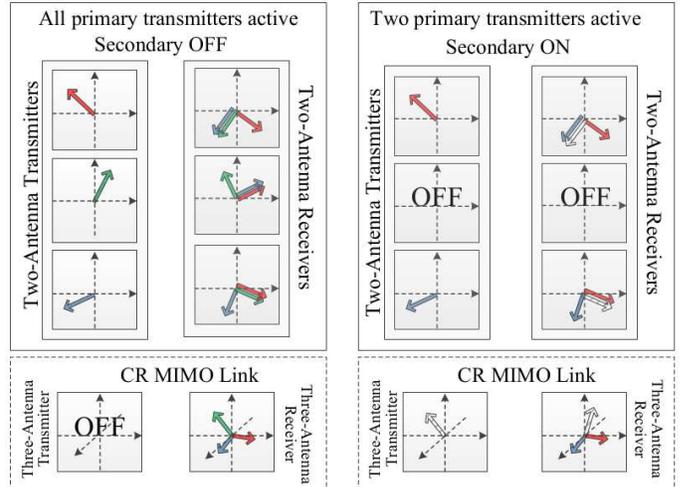}
\vspace{-1cm}
\caption{A secondary system in the presence of a primary 3-user IA system; left: all the primary transmitters are active and there is no unused DoFs. right: the second primary transmitter is off and there is one unused DoF that can be used by the secondary system.}
\vspace{-0.5cm}
\label{vector}
\end{figure}
%%%%%%%%%%%%%%%%%%%%%%%%%%%%%%%%%%%%%%%%%%%%%%%%%%%%%%%%%
To realize the interference-free channels from the secondary transmitter to the primary receivers, one can design the secondary precoder matrix to align the interference in each primary receiver separately as follows:
\begin{equation}
\left[ {\begin{array}{*{20}{c}}
\textbf{U}^{[1]\dag}\textbf{H}^{[10]}\\
\textbf{U}^{[2]\dag}\textbf{H}^{[20]}\\
 \vdots \\
\textbf{U}^{[K]\dag}\textbf{H}^{[K0]}
\end{array}} \right]\textbf{V}^{[0]}=\textbf{P}\textbf{V}^{[0]}=\textbf{0}_{KM^{[0]} \times d^{[0]}}.
\end{equation}
The size of the $i^{th}$ matrix in $\textbf{P}$ is $d^{[i]}\times M^{[0]}$, and therefore, when all streams of the primary transmitters are active, the rank of $\textbf{P}$ can be written as:
\begin{equation}
\text{rank}(\textbf{P})=\min(\sum_{i=1}^{K}d^{[i]},M^{[0]}).
\end{equation}
By denoting the number of active streams in the $i^{th}$ primary transmitter as $d_\mathcal{A}^{[i]}$, the rank of $\textbf{P}$ becomes:
\begin{equation}
\text{rank}(\textbf{P})=\min(\sum_{i =1}^{K}d_\mathcal{A}^{[i]},M^{[0]}),
\end{equation}
and if $\sum_{i =1}^{K}d_\mathcal{A}^{[i]}< M^{[0]}$, then $\min(\sum_{i =1}^{K}d_\mathcal{A}^{[i]},M^{[0]})=\sum_{i =1}^{K}d_\mathcal{A}^{[i]}$. So, one can simply find the dimension of the null space of $\textbf{P}$ as:
\begin{equation}
\text{null}\left(\textbf{P}\right)=M^{[0]}-\sum_{i =1}^{K}d_\mathcal{A}^{[i]}.
\end{equation}
Therefore, the number of transmit antennas and data streams at the secondary should satisfy the following equation:
\begin{equation}
\label{eq21}
\text{null}\left(\textbf{P}\right)=M^{[0]}-\sum_{i =1}^{K}d_\mathcal{A}^{[i]} \geq d^{[0]},
\end{equation}
To have zero interference at all the primary receivers, we should then have (\ref{eq:lemma1}).
\end{IEEEproof}

\newtheorem{remark}{Remark}
\begin{remark} The inequality \eqref{eq:lemma1} of \textit{Lemma 1} can be considered as the maximum number of streams that the secondary system is able to transmit which is limited by $d^{[0]} \leq M^{[0]}-\sum_{i =1}^{K}d_\mathcal{A}^{[i]}$. More precisely, if we assume the same number of secondary transmit antennas as the DoFs of the primary system, i.e., $\sum_{i=1}^{K}d^{[i]}$, the CR network is able to use all the unused DoFs of the primary.
\end{remark}

The $c^{th}$ column of the precoder matrix $\textbf{V}^{[0]}$ ($c=1,\cdots,d^{[0]}$) can be defined by the eigenvector corresponding to one of the $\sum_{i =1}^{K}\left(d^{[i]}-d_\mathcal{A}^{[i]}\right)$ zero eigenvalues of $\bf{P}$; i.e.:
\begin{equation}
\textbf{V}^{[0]}_{\parallel c}=\text{E}_{l}[\textbf{P}]
\end{equation}
where $l=\sum_{i =1}^{K}d_\mathcal{A}^{[i]}+c$.

\textit{Example:} As shown in Fig.~\ref{vector}, consider a $3$-user MIMO IA primary system in which all the nodes have two antennas, and each primary transmitter sends one stream of data to its corresponding receiver. We have also assumed a $3\times 3$ secondary link. If all the primary transmitters are active, there will be no unused DoFs to be used by the secondary. However, when only two or less primary transmitters are active, knowing that the number of transmit antennas at the secondary satisfy the condition in (\ref{eq:lemma1}), the CR network is able to use the unused DoF(s) of the primary system. For the case of one unused DoF, the secondary transmitter can select the subspace corresponding to the null space of the secondary to primary interference channel, i.e., $\textbf{V}^{[0]}_{\parallel 1}=\text{E}_{3}[\textbf{P}]$ as its precoder.

\begin{remark}[Underlay CR Approach] If the null space of $\bf{P}$ is empty, no secondary precoder exists that nullifies the interference at all the primary receivers. In such a case, one can consider an underlay CR approach in which the interference to all the primary receivers can be kept below a predefined threshold $\eta$; i.e.:
\begin{equation}\label{eq:underlay}
I^{[1]}_{\text{CR}} \simeq I^{[2]}_{\text{CR}} \simeq  ... \simeq I^{[K]}_{\text{CR}} \leq \eta.
\end{equation}
In this paper, we do not consider such an underlay approach, and assume that there are some unused DoFs in the primary system that can be utilized by the secondary. 
However, without considering the individual constraint on the interference leakage in \eqref{eq:underlay}, the interference minimization problem \eqref{precoder} can still be considered while the number of transmit antennas of the secondary network does not satisfy the condition in \textit{Lemma 1}. In this case, considering $\text{Tr}\left[\textbf{A}+\textbf{B}\right]=\text{Tr}\left[\textbf{A}\right]+\text{Tr}\left[\textbf{B}\right]$ and $\text{Tr}\left[\textbf{A}\textbf{B}\right]=\text{Tr}\left[\textbf{B}\textbf{A}\right]$, we can rewrite the objective function of \eqref{precoder} as:
\begin{align}\label{opt1}
&\min_{\textbf{V}^{[0]}} \text{Tr}\left[\textbf{V}^{[0]\dag}\textbf{Q}\textbf{V}^{[0]}\right] \\ \notag
&\text{s.t.~} \textbf{V}^{[0]\dag}\textbf{V}^{[0]}=\textbf{I}_{d^{[0]}},
\end{align}
where $\textbf{Q}$ is an $M^{[0]} \times M^{[0]}$ matrix such that:
\begin{equation}
\textbf{Q}=\sum_{k=1}^{K}\frac{p^{[0]}}{d^{[0]}}\textbf{H}^{[k0]\dag}\textbf{U}^{[k]}\textbf{U}^{[k]\dag}\textbf{H}^{[k0]}.
\end{equation}
Since $\textbf{Q}$ is Hermitian, by using trace minimization \cite[p.191]{Horn}, the total interference leakage is the summation of $d^{[0]}$ smallest eigenvalues of $\textbf{Q}$.
\end{remark}

\subsection{Decoder at Secondary Receiver}
% \subsubsection{SINR maximization}
The principal goal of the CR network is to agilely utilize the unused DoFs in such a way to optimize a performance objective. In this paper, we consider the secondary SINR maximization approach to design the secondary decoder. 

\begin{lemma}[Decoder matrix] The $l^{th}$ column of the decoder matrix $\textbf{U}^{[0]}$ to maximize the SINR can be derived as:
\begin{equation}\label{eq:25}
\textbf{U}^{[0]}_{\parallel l}=\dfrac{\left(\textbf{B}_{l}\right)^{-1}\textbf{H}^{[00]}\textbf{V}^{[0]}_{\parallel l}}{||\left(\textbf{B}_{l}\right)^{-1}\textbf{H}^{[00]}\textbf{V}^{[0]}_{\parallel l}||},
\end{equation}
and the maximum SINR achieved by this solution is:
\begin{equation} \label{SINR}
\gamma^{\rm CR max}_{l}=\dfrac{p^{[0]}}{d^{[0]}}\textbf{V}^{[0] \dag}_{\parallel l}\textbf{H}^{[00]\dag}\left(\textbf{B}_{l}\right)^{-1}\textbf{H}^{[00]}\textbf{V}^{[0]}_{\parallel l}.
\end{equation}
\end{lemma}
\begin{IEEEproof}
The proof is in \cite[Section V.C]{Gomadam2011}.
\end{IEEEproof}
%{\color{red}Size of receive antenna?}
\begin{remark}If we define the normalized vector $H_l=\textbf{H}^{[00]}\textbf{V}^{[0]}_{\parallel l}$, the instantaneous SINR at the CR receiver can be bounded by the Kantorovich matrix inequality \cite{Baksalary1991}:\footnotesize
\begin{equation}
\left(H_l^{\dag}\textbf{B}_{l}H_{l}\right)^{-1}\leq \gamma^{\rm CR max}_{l}  \leq \frac{(\lambda_1^l+\lambda_{N^{[0]}}^l)^2 }{4\lambda_1^l \lambda_{N^{[0]}}^l}\left(H_l^{\dag}\textbf{B}_{l}H_l\right)^{-1},
\end{equation}\normalsize
where $\lambda_1^l$ and $\lambda_{N^{[0]}}^l$ are the largest and the smallest eigenvalues of $\textbf{B}_{l}$, respectively.
\end{remark}

\begin{lemma}[Number of receive antennas] Assuming single stream transmission in the CR network, the average SINR at the CR receiver is an increasing function of the number of receive antennas at the secondary.
\end{lemma}
\begin{IEEEproof}
The proof is in Appendix \ref{app2}.
\end{IEEEproof}
Even though \textit{Lemma 3} concerns with the case of single data stream at the secondary network, its statement is valid for the case of transmission of more than one stream. We have shown this with numerical simulations in Section IV.

\textit{Example:} As depicted in Fig.~\ref{vector}, the received vectors at the secondary receiver from two active primary transmitters only span two dimensions out of three available dimensions of receiver. At high SNRs, the optimal decoder is the vector perpendicular to the two-dimensional plane which is spanned by two interference vectors from active primary transmitters (see equation \eqref{eq:25}). By increasing the number of antennas of the secondary receiver, there will be more degrees of freedom to choose the optimal decoder. This helps in increasing the received SINR of the secondary as stated in {\it{Lemma 3}}.

%%%%%%%%%%%%%%%%%%%%%%%%%%%%%%%%%%%%%%%%%%%%%%%%%%%%%%%%%%%%%
\section{Null Space Sensing and Detection}
%%%%%%%%%%%%%%%%%%%%%%%%%%%%%%%%%%%%%%%%%%%%%%%%%%%%%%%%%%%%%
In the previous section, we showed that the unused DoFs of the primary network can be utilized by  the secondary MIMO system. In this section, we provide a fast sensing technique to identify the availability of null space of the primary. In the proposed model, the CR receiver has the role of a fusion center and has to detect the availability of spatial holes (null spaces) of the concurrent primary $K$-user IA system. In this scheme, a spectrum sensing method  at the CR fusion center determines the availability of the spatial holes. Note that using this scheme, the fusion center cannot identify the indices of the unused DoFs, or equivalently, inactive data streams. The main reason for applying this method is to decrease the computational load in fusion center to sense the availability of inactive DoFs. To find the indices of the inactive primary streams, a search method will be introduced in the next section. 

The null space sensing problem can be formulated as a hypothesis test. In the proposed hypothesis test, $\mathcal{H}_0$ implies that there exists an inactive data stream and $\mathcal{H}_1$ indicates that the primary null space is empty. Considering the set of received vectors from the primary transmitters, we can rewrite the hypotheses as:
\begin{align}
&\mathcal{H}_0: Y[n] =\mathfrak{X}_0[n]+Z[n], \;\;n=0, ..., L-1, \\ \notag
&\mathcal{H}_1: Y[n] = \mathfrak{X}_1[n]+Z[n], \;\;n=0, ..., L-1, \\ \notag
\end{align}
where:
\begin{equation}\label{27}
\mathfrak{X}_0[n]=\mathop{\sum_{k=1}^{K}}\mathop{\sum^{d^{[k]}}_{j=1}}_{j\neq l^{[i]}}\textbf{H}^{[0k]}\textbf{V}^{[k]}_{\parallel j}\bar{X}^{[k]}_j[n],
\end{equation}
when the $l^{th}$ stream of the $i^{th}$ user is inactive, and:
\begin{equation}
\mathfrak{X}_1[n]=\sum_{k=1}^{K} \sum^{d^{[k]}}_{j=1}\textbf{H}^{[0k]}\textbf{V}^{[k]}_{\parallel j}\bar{X}^{[k]}_j[n],
\end{equation}
in the case that all the streams are active. $Z[n]$ is the zero mean Gaussian noise vector at the fusion center. For estimating the covariance matrix of the received signal, we firstly define the \emph{smoothing factor} $T$ such that $T \geq N^{[0]}$ \cite{Zeng2009a}. By considering the received and noise vectors over $T$ consecutive sample times, we can define the following matrices:
\begin{align}
&\bar{\textbf{Y}}[n]\stackrel{\text{def}}{=}\left[Y^T[n],Y^T[n-1], \cdots,Y^T[n-T+1]\right]^T, \\ \notag
&\bar{\mathfrak{{X}}_i}[n]\stackrel{\text{def}}{=}\left[\mathfrak{X}_i^T[n],\mathfrak{X}_i^T[n-1], \cdots,\mathfrak{X}_i^T[n-T+1]\right]^T
, i=0,1, \\ \notag
&\bar{\textbf{Z}}[n]\stackrel{\text{def}}{=}\left[Z^T[n],Z^T[n-1], \cdots,Z^T[n-T+1]\right]^T.
\end{align}
Assuming that there are a total of $L$ such matrices, the covariance matrices normalized by the known noise variance for the $L$ collected sampled matrices can be written as:
\begin{align}
&\textbf{R}_{Y}=\frac{1}{\sigma_z^2}\E\{\bar{\textbf{Y}}[n]\bar{\textbf{Y}}^{\dag}[n]\}\approx \frac{1}{L\sigma_z^2}\sum_{n=T-1}^{T+L-2}\bar{\textbf{Y}}[n]\bar{\textbf{Y}}^{\dag}[n],  \\ \label{eq:30}
&\textbf{R}_{\mathfrak{X}_i}=\frac{1}{\sigma_z^2}\E\{\bar{\mathfrak{X}_i}[n]\bar{\mathfrak{X}}_i^{\dag}[n]\}\approx\frac{1}{L\sigma_z^2}\sum_{n=T-1}^{T+L-2}\bar{\mathfrak{X}_0}[n]\bar{\mathfrak{X}_0}^{\dag}[n],\\
&\textbf{R}_{Z}=\frac{1}{\sigma_z^2}\E\{\bar{\textbf{Z}}[n]\bar{\textbf{Z}}^{\dag}[n]\}\approx\frac{1}{L\sigma_z^2}\sum_{n=T-1}^{T+L-2}\bar{\textbf{Z}}[n]\bar{\textbf{Z}}^{\dag}[n].
\label{eq:32}
\end{align}
Since the noise and data are independent, the received covariance matrix in hypothesis $\mathcal{H}_0$ can be written as:
\begin{equation}
\textbf{R}_{Y}=\textbf{R}_{\mathfrak{X}_0}+\textbf{R}_{Z}.
\end{equation}

Therefore, the received covariance matrix is the summation of two covariance matrices. If we assume that the secondary has the same number of antennas as the total DoFs of the primary network, it can be simply shown that the $N^{[0]th}$ (smallest) eigenvalue of $\textbf{R}_{\mathfrak{X}_0}$ is zero. We can employ the eigenvalue analysis methods for sensing the availability of null space. Using properties of eigenvalues for spectrum sensing has been studied for other networks \cite{Zeng2009,Abreu2011,Wei2010}. Here, we propose an eigenvalue based sensing method for a $K-$user IA system as a novel application of this sensing method. 

\begin{lemma}[Fast Eigenvalue Sensing]
The probability of a false alarm (PFA) for a predefined threshold $\eta$ is bounded as:
\begin{equation}\label{ineq_new}
{\rm{Pr}}(\lambda_{min}(\textbf{R}_{Z}) > \eta) \leq {P}_{FA} \leq {\rm{Pr}}(\lambda_{max}(\textbf{R}_{Z}) > \eta),
\end{equation}
where ${P}_{FA}={\rm{Pr}}\left(\lambda_{min}(\textbf{R}_{Y}) > \eta|\mathcal{H}_0\right)$ and $N^{[0]}=\sum_{i=1}^{K}{d^{[i]}}$.
\end{lemma}

\begin{IEEEproof}
The Weyl's inequality for the summation of eigenvalues in hypothesis $\mathcal{H}_0$ states that:
\begin{small}
\begin{equation}\label{ineq1}
\lambda_{min}(\textbf{R}_{\mathfrak{X}_0})+\lambda_{min}(\textbf{R}_{Z}) \leq \lambda_{min}(\textbf{R}_{Y}) \leq \lambda_{min}(\textbf{R}_{\mathfrak{X}_0})+\lambda_{max}(\textbf{R}_{Z})
\end{equation}
\end{small}
If we assume that the number of antennas at the fusion center is equal to the total number of data streams of the $K$-user IA system ($N^{[0]}=\sum_{i=1}^{K}d^{[i]}$), the smallest eigenvalue of the covariance matrix, $\textbf{R}_{\mathfrak{X}_0}$, shows the presence or absence of a spatial hole. Then, the inequality \eqref{ineq1} can be changed to:
\begin{equation}
\lambda_{min}(\textbf{R}_{Z}) \leq \lambda_{min}(\textbf{R}_{Y}) \leq \lambda_{max}(\textbf{R}_{Z})
\end{equation}
Note that $\textbf{R}_{Z}$ is statistically known at CR fusion center. In such a case, the PFA for the a predefined threshold $\eta$ is bounded as:
\begin{equation}
{\rm{Pr}}(\lambda_{min}(\textbf{R}_{Z}) > \eta) \leq {P}_{FA} \leq {\rm{Pr}}(\lambda_{max}(\textbf{R}_{Z}) > \eta)
\end{equation}
where ${P}_{FA}={\rm{Pr}}\left(\lambda_{min}(\textbf{R}_{Y}) > \eta|\mathcal{H}_0\right)$.
\end{IEEEproof}

The noise vectors are $N^{[0]}$-variate complex Gaussian vectors with covariance matrix ${\bf{\Sigma}}=\sigma_z^2\textbf{I}$ and are assumed to be independent. Therefore, the covariance matrix $\textbf{R}_{Z}$, as calculated in (\ref{eq:32}), has a central complex Wishart distribution with $LT$ degrees of freedom and covariance matrix ${\bf{\Sigma}}$, i.e., $\textbf{R}_{Z} \sim \mathcal{W}_{N^{[0]}}\left( LT,{\bf{\Sigma}} \right)$ \cite{verduRMT}. Note that the distribution of random matrix $\textbf{R}_{Y}$ is not available. However, according to (\ref{ineq_new}), the threshold $\eta$ can be set properly by knowing the distribution of the smallest and largest eigenvalues of $\textbf{R}_{Z}$. Since $\textbf{R}_{Z}$ has a Wishart distribution, its covariance matrix ${\bf{\Sigma}}$ is a sufficient statistic for setting the threshold $\eta$ \cite[Chap. 7]{Anderson2003}.

As explained, the proposed sensing method only detects the presence or absence of unused DoFs. Therefore, this scheme only serves a coarse detection step. The detection of the exact number of inactive data streams and their indices is provided in the next section. Note that if the coarse sensing step cannot make the correct decision about the presence of spatial hole, the fine detection in next step will not be performed which translates in a performance degradation of the total sensing scheme. Hence, the PFA plays a crucial role in the sensing. We, next, consider the upper bound on the PFA to find a suitable threshold $\eta$. 

The cumulative distribution function (CDF) of the largest or smallest eigenvalue of a Wishart matrix is a well-known problem in random matrix theory \cite{verduRMT}. For the case of central Wishart matrices, Khatri's result provides the closed-form function as \cite{C.G.Khatri2013,Kang2003}:
\begin{equation}
{\rm{Pr}}(\lambda_{max}(\textbf{R}_{Z}) \leq \eta)=\frac{|\boldsymbol\Psi \left(\eta\right)|}{\prod_{k=1}^{N^{[0]}}\Gamma\left(LT-k+1\right)\left(N^{[0]}-k+1\right)}
\end{equation}
where $|\cdot|$ denotes the determinant, and $\boldsymbol\Psi \left(\eta\right)$ is an $N^{[0]}\times N^{[0]}$ Henkel matrix function of $\eta \in (0,\infty)$ with entries given by
\begin{equation}
\{\boldsymbol\Psi \left(\eta\right)\}_{i,j}=\Gamma\left(LT-N^{[0]}+i+j-1,L\eta\right), i=1,\cdots,N^{[0]}
\end{equation}
where $\Gamma(\cdot,\cdot)$ is the incomplete Gamma function. While Khatri's formulation provides the closed form CDF, it is difficult to use it to derive a closed-form easy-to-use equation for $\eta$. Therefore, an approximation method for deriving the distribution of the largest and smallest eigenvalues has been recently proposed in literature \cite{Ma2012}. When $\lim_{LT \rightarrow \infty}\frac{N^{[0]}}{LT}=y$ where $0<y<1$, the CDFs of the largest and smallest eigenvalues approximately converge to the Tracy-Widom distribution of order two; i.e. for the largest eigenvalue:
\begin{align}
&{\rm{Pr}}(\lambda_{max}(\textbf{R}_{Z}) \leq \eta) \approx \\ \notag
&\;\;\;\;\;\; {F_2}\left(\frac{L\eta-\left(\sqrt{LT}+\sqrt{N^{[0]}}\right)^2}{\left(\sqrt{LT}+\sqrt{N^{[0]}}\right)\left(\sqrt{\frac{1}{LT}}+\sqrt{\frac{1}{N^{[0]}}}\right)^{\frac{1}{3}}}\right)
\end{align}
and for the smallest eigenvalue:
\begin{align}
& {\rm{Pr}}(\lambda_{min}(\textbf{R}_{Z}) \leq \eta)\approx \\ \notag
&\;\;\;\;\;\; {F}_2\left(\frac{L\eta-\left(\sqrt{LT}-\sqrt{N^{[0]}}\right)^2}{\left(\sqrt{LT}-\sqrt{N^{[0]}}\right)\left(\sqrt{\frac{1}{LT}}-\sqrt{\frac{1}{N^{[0]}}}\right)^{\frac{1}{3}}}\right)
\end{align}
where $F_2$ is the Tracy-Widom distribution of order two and can be written as:
\begin{equation}
{F}_2\left(\lambda\right)=\exp\left(-\int\limits_\lambda ^\infty  {\left(x - \lambda \right)q^2(x)dx}\right),
\end{equation}
where $q(x)$ is the solution to the non-linear Painlev's equation of type II, i.e.:
\begin{equation}
q''(x)=xq(x)+2q^3(x).
\end{equation}
Using this approximate method, and assuming the availability of the inverse of Tracy-Widom distribution function, the threshold value can be derived as:
\small
\begin{align}
\eta&=\frac{1}{L} \left(\sqrt{LT}+\sqrt{N^{[0]}}\right)\left(\sqrt{\frac{1}{LT}}+\sqrt{\frac{1}{N^{[0]}}}\right)^{\frac{1}{3}}{F}_2^{-1}(1- {P}_{FA}) \\ \notag
& + \frac{1}{L}\left(\sqrt{LT}+\sqrt{N^{[0]}}\right)^2.
\end{align}
\normalsize
Thus, if we consider the minimum eigenvalue of the received covariance matrix in \eqref{ineq1} as the test statistic, since the degree of freedom of the Wishart distribution in the received covariance matrix is the multiplication of two sampling parameters, $T$ and $L$, the PFA has sharp decreasing behaviour. By proper selection of the threshold value based on a reasonable PFA, this method provides a fast scanning scheme while keeping the probability of correct detection of a spatial hole high.

It is worth mentioning that, for the sake of simplicity, we considered the hypothesis $\mathcal{H}_0$ in \eqref{27} as the case when there is only one inactive primary stream. The proposed sensing method, however, is general and includes the case of the detection of more than one spatial hole. In this case, the minimum eigenvalue of covariance matrix of the signal, $\textbf{R}_{\mathfrak{X}_0}$, in \eqref{ineq1} is still zero. We will show in the simulation results that in this case the PFA can be slightly lower than the case of only one inactive stream.
% On the other hand, when all users are active, $\lambda_{min}(\textbf{R}_{\mathfrak{X}_1})$ has non-zero value and it is not changed significantly by higher $LT$ values. 
%%%%%%%%%%%%%%%%%%%%%%%%%%%%%%%%%%%%%%%%%%%%%%%%%%%%%%%%%%%%%
\section{Search for the Index Set of Unused DoFs}
%%%%%%%%%%%%%%%%%%%%%%%%%%%%%%%%%%%%%%%%%%%%%%%%%%%%%%%%%%%%%
Since the interference channels of the CR receiver ($\textbf{H}^{[0i]}, i=1,...,K$) are not aligned, the transformations applied by the precoder vectors of primary transmitters do not change the independent nature of the received signal vectors at this receiver. Therefore, the CR fusion center can sense the presence or absence of each stream of primary by an independent binary hypothesis test \cite{Zeng2010,Axell}. To apply hypothesis testing to find out the indices of the unused DoFs, we assume that the secondary receiver first applies a sensing vector to the received signal. 
\subsection{Secondary Sensing Vectors}
The second phase of the proposed sensing scheme is applied to the received signal at the fusion center when the presence of spatial hole is detected by the aforementioned eigenvalue based sensing scheme. In this phase, the fusion center scans the presence of each DoF of primary network by searching towards the appropriate directions of the received vectors. This search method can be accomplished by finding a sensing vector set consisting of $\sum_{i=1}^{K}d^{[i]}$ vectors $D_l^{[i]}$ ($i=1,\cdots,K$; $l=1,\cdots,d^{[i]}$). To avoid the effect of other streams in the sensing of the $j^{th}$ stream, the direction of the $j^{th}$ sensing vector is selected such that it is orthogonal to the summation of all the other received streams except the $j^{th}$ stream. Thus, the sensing vector $D_l^{[i]}$ for finding the presence of the $l^{th}$ stream of the $i^{th}$ primary user is defined as:
% For simplification in notations, let us define the received matrix from primary user $i^{th}$ at the CR fusion center as $\textbf{h}_i=\textbf{H}^{[0i]}\textbf{V}^{[i]}$ the 
\begin{equation} \label{ortho}
D_l^{[i]} \perp \mathop{\sum_{k=1}^{K}}\mathop{\sum^{d^{[k]}}_{j=1}}_{j\neq l^{[i]}}\textbf{H}^{[0k]}\textbf{V}^{[k]}_{\parallel j},
\end{equation}
for $i=1,\cdots,K$ and $l=1,\cdots,d^{[i]}$, where the sensing vector $D_l^{[i]}$ is of size $N^{[0]} \times 1$ and $D_l^{[i]\dag}D_l^{[i]}=1$. By this selection, $D_l^{[i]}$  is orthogonal to the subspace spanned by all the received vectors except the one corresponding to the $l^{th}$ stream of the $i^{th}$ primary user. In other words, we can find the vector $D_l^{[i]}$ such that its inner product with the corresponding received vector is maximized, while $D_l^{[i]} \notin \text{span}\{\textbf{H}^{[0k]}\textbf{V}^{[k]}_{\parallel j}\}$ for $k=1,...,K$, $k \neq i$ and $j=1,...,d^{[k]}$, $j\neq l$. By rewriting \eqref{ortho} as an inner product, we can find the optimal direction of each sensing vector by formulating the following optimization problem:
\begin{align}\label{optr}
\max_{D_l^{[i]}}&\;\;D_l^{[i]\dag}\textbf{H}^{[0i]}\textbf{V}^{[i]}_{\parallel l} \\ \notag
\text{s.t. } &D_l^{[i] \dag}\textbf{R}_l^{[i]}=\textbf{0}_{1\times (\sum_{i=1}^{{K}}{d^{[i]}-1})}, \\ \notag
&D_l^{[i]\dag}D_l^{[i]}=1,
\end{align}
where the matrix $\textbf{R}_l^{[i]}$ of size $N^{[0]}\times (\sum_{i=1}^{{K}}{d^{[i]}-1})$  is:
\begin{equation}
\textbf{R}_l^{[i]}=\left[\textbf{H}^{[01]}\textbf{V}^{[1]}, ... , \textbf{H}^{[0i]}\textbf{V}^{[i]}_{\parallel l-1}, \textbf{H}^{[0i]}\textbf{V}^{[i]}_{\parallel l+1}  , ... , \textbf{H}^{[0K]}\textbf{V}^{[K]} \right].
\end{equation}

\begin{lemma}
The sensing vector set is the solution of the optimization problem \eqref{optr} and can be written as:
\begin{equation}\label{eq:sensvec}
D_l^{[i]}=\frac{1}{\lambda}\left(\textbf{H}^{[0i]}\textbf{V}^{[i]}_{\parallel l}-C_l^{[i]}\right),
\end{equation}
where
\begin{equation}\label{R}
C_l^{[i]}=\textbf{R}_l^{[i]}\left(\textbf{R}_l^{[i]\dag}\textbf{R}_l^{[i]}\right)^{-1}\textbf{R}_l^{[i]\dag}\textbf{H}^{[0i]}\textbf{V}^{[i]}_{\parallel l} ,
\end{equation}
and
\begin{equation}
\lambda^2=\left(\textbf{V}^{[i]\dag}_{\parallel l}\textbf{H}^{[0i]\dag}-C_l^{[i]\dag}\right)\left(\textbf{H}^{[0i]}\textbf{V}^{[i]}_{\parallel l}-C_l^{[i]}\right).
\end{equation}
\end{lemma}
\begin{IEEEproof}
The proof is in Appendix B.
\end{IEEEproof}

\begin{remark}[Feasibility of Sensing]
The minimum required number of antennas in the fusion center is equal to the total active  DoFs of the primary system. In fact, the rank of matrix $\textbf{R}_l^{[i]}$ must be less than the number of secondary receive antennas,  $N^{[0]}$. We should have $N^{[0]} > \sum_{i=1}^{{K}}{d^{[i]}-1}$ antennas to find a vector in null space of $\textbf{R}_l^{[i]}$ to maximize the inner product in the direction of the desired vector, $\textbf{H}^{[0i]}\textbf{V}^{[i]}$.
\end{remark}

\subsection{Sensing Criterion and the Probability of False Alarm}
Based on the available set of sensing vectors, the fusion center is able to perform the scanning by multiplying the received signal by each sensing vector. If we consider $T$ samples of the received signal, the hypothesis testing problem for the $l^{th}$ stream of the $i^{th}$ primary user can be written as:
\begin{align}
&\mathcal{H}_0: {y}^{[i]}_l[n] =\tilde{z}[n], \;\;n=0, ..., T-1\\ \notag 
&\mathcal{H}_1: {y}^{[i]}_l[n] = \frac{p^{[i]}}{d^{[i]}}D_{l}^{[i] \dag}\textbf{H}^{[0i]}\textbf{V}^{[i]}_{\parallel l}\bar{X}^{[i]}_l[n]+\tilde{z}[n], \;\;n=0, ..., T-1
\end{align}
where $\tilde{z}[n]$ is the output noise after the decoder. Here, since we consider normalized sensing vectors in the second constraint of the optimization problem \eqref{optr}, the statistics of the noise $\tilde{z}[n]$ after multiplication remain unchanged.

In this scenario, there is an uncertainty about the amplitude of the signal in $\mathcal{H}_1$ hypothesis because of the unknown power allocation scheme in the primary transmitters. In classical detection theory, there are two main approaches to tackle this problem: the Bayesian and the GLRT methods \cite{Zeng2010,Axell}. In the Bayesian method, the unknown parameter can be treated as a random variable with known distribution \cite{Zeng2010}. The likelihood functions can then be achieved by calculating the marginal probability based on the prior distribution of the unknown parameter. Since the choice of {\textit{a priori}} distributions affects the detection performance dramatically and also calculating the marginal distributions is often not tractable, in this paper we adopt the GLRT as a suboptimal detection method \cite[Sec. 6.4.2]{StevenM.Kay1998}.

In this method, we first estimate the unknown parameters, and then use a likelihood ratio test to make a decision. We assume that the distribution of the signal is complex Gaussian with zero mean and unknown variance. So, if we consider the signal term in $\mathcal{H}_1$, i.e.:
\begin{equation}
s^{[i]}_l[n]=\frac{p^{[i]}}{d^{[i]}}D_{l}^{[i] \dag}\textbf{H}^{[0i]}\textbf{V}^{[i]}_{\parallel l}\bar{X}^{[i]}_l[n],
\end{equation}
and by converting into vector representation, the hypothesis test can be rewritten as:
\begin{align}
&\mathcal{H}_0: Y^{[i]}_l =\tilde{Z}, \\ \notag
&\mathcal{H}_1: Y^{[i]}_l = S^{[i]}_{ l }+\tilde{Z},
\end{align}
where the sample vectors $Y^{[i]}_l$, $S^{[i]}_{l}$, and $\tilde{Z}$ are of size $T \times 1$.

The general solution for such a hypothesis testing with unknown parameter is to estimate the unknown parameter by the well-known maximum likelihood estimation (MLE) method under $\mathcal{H}_1$, i.e.:
\begin{equation}
\hat{\Theta}_1=\mathop{\text{argmax}}_{\Theta_1}\mathit{p}(Y^{[i]}_l|\mathcal{H}_1,\Theta_1),
\end{equation}
where $\mathit{p}(\cdot)$ denotes the probability density function (PDF). Note again that our assumption is that we have no knowledge of the signal variance, but the noise variance $\sigma^2_z$ is assumed known, i.e. $\tilde{Z}\sim \mathcal{C}\mathcal{N}(\textbf{0}_{T \times 1},\sigma_z^2\textbf{I}_T)$. It is also assumed that $S^{[i]}_{ l }$ and $\tilde{Z}$ are independent and jointly Gaussian. 

Based on Neyman-Pearson (NP) theorem \cite[Sec. 3.3]{StevenM.Kay1998}, for a given PFA, the test statistic that
maximizes the probability of detection (PD) is:
\begin{equation}\label{GLRT1}
L_{\text{GLRT}}=\frac{\mathit{p}(Y^{[i]}_l|\mathcal{H}_1,\hat{\Theta}_1)}{\mathit{p}(Y^{[i]}_l|\mathcal{H}_0)} \mathop{\gtrless}_{\mathcal{H}_0}^{\mathcal{H}_1} \eta,
\end{equation}
where there is no unknown parameter in hypothesis $\mathcal{H}_0$, and we have:
\begin{equation}\label{H0}
\mathit{p}(Y^{[i]}_l|\mathcal{H}_0)=\prod_{n=0}^{T-1}\frac{1}{2\pi\sigma_z^2}\text{exp}\left[-\frac{1}{2\sigma^2_z}||{y}^{[i]}_l[n]||^2\right]
\end{equation}
Hence:
\begin{equation}\label{H0log}
\ln\mathit{p}(Y^{[i]}_l|\mathcal{H}_0)=-T\ln\left(2\pi\sigma_z^2\right)-\frac{1}{2\sigma^2_z}Y^{[i]\dag}_{l}Y^{[i]}_l.
\end{equation}
On the other hand, the likelihood function under $\mathcal{H}_1$ with the unknown parameter $\sigma_s$ can be written as:
\small
\begin{equation}
\mathit{p}(Y^{[i]}_l|\mathcal{H}_1,\sigma_s^2)=\prod_{n=0}^{T-1}\frac{1}{2\pi(\sigma^2_s+\sigma_z^2)}\text{exp}\left[-\frac{1}{2(\sigma^2_s+\sigma_z^2)}||{y}^{[i]}_l[n]||^2\right].
\end{equation}
\normalsize
Taking logarithm from both sides, we get:
\begin{equation}\label{loglik}
\ln\mathit{p}(Y^{[i]}_l|\mathcal{H}_1,\sigma_s^2)=-T\ln\left(2\pi(\sigma^2_s+\sigma_z^2)\right)-\frac{1}{2(\sigma^2_s+\sigma_z^2)}Y^{[i]\dag}_{l}Y^{[i]}_l.
\end{equation}
The first derivative of the log-likelihood function with respect to $\sigma_s^2$ is:
\begin{equation}
\frac{\partial\ln\mathit{p}(Y^{[i]}_l|\mathcal{H}_1,\sigma_s^2)}{\partial\sigma_s^2}=-\frac{T}{\sigma_s^2+\sigma_z^2}+\frac{1}{2(\sigma^2_s+\sigma_z^2)^2}Y^{[i]\dag}_{l}Y^{[i]}_l.
\end{equation}
Therefore, the MLE of the unknown parameter $\sigma_s^2$ can be obtained by finding the root of the first derivative as:
\begin{equation}\label{estimated}
\hat{\sigma}_s^2=\frac{Y^{[i]\dag}_{l}Y^{[i]}_l}{2T}-\sigma_z^2.
\end{equation}
By substituting \eqref{estimated} in \eqref{loglik}, we have:
\begin{equation}\label{estloglik}
\ln\mathit{p}(Y^{[i]}_l|\mathcal{H}_1,\hat{\sigma}_s^2)=-T\ln\left(\frac{\pi Y^{[i]\dag}_{l}Y^{[i]}_l}{T}\right)-T.
\end{equation}
By plugging \eqref{H0log} and \eqref{estloglik} into \eqref{GLRT1}, the log-GLRT statistics becomes:
\begin{equation}
\ln L_{\text{GLRT}}(Y^{[i]}_l)=T\ln\left(\frac{2T\sigma_z^2}{Y^{[i]\dag}_{l}Y^{[i]}_l}\right)+\frac{1}{2\sigma^2_z}Y^{[i]\dag}_{l}Y^{[i]}_l-T,
\end{equation}
Therefore, the final test can be written as:
\begin{equation}
L_{\text{GLRT}}(Y^{[i]}_l)=\left(\frac{2T\sigma_z^2}{Y^{[i]\dag}_{l}Y^{[i]}_l}\right)^T\exp\left(\frac{1}{2\sigma^2_z}Y^{[i]\dag}_{l}Y^{[i]}_l-T \right)\mathop{\gtrless}_{\mathcal{H}_0}^{\mathcal{H}_1} \eta.
\end{equation}
We can define the new threshold for the test statistics as:
\begin{equation}\label{Threshold}
\mathcal{T}(Y^{[i]}_l)=\frac{1}{Y^{[i]\dag}_{l}Y^{[i]}_l}\exp\left(\frac{Y^{[i]\dag}_{l}Y^{[i]}_l}{2T\sigma^2_z} \right)\mathop{\gtrless}_{\mathcal{H}_0}^{\mathcal{H}_1} \frac{\eta^{\frac{1}{T}}}{2T\sigma_z^2}e^1=\eta'.
\end{equation}
Similar to the NP theorem \cite[Sec. 3.3]{StevenM.Kay1998}, the threshold $\eta'$ can be obtained for an arbitrary PFA, i.e.:
\begin{equation}
{P}_{\rm FA}=\mathit{p}(\mathcal{T}(Y^{[i]}_l)>\eta' ;\mathcal{H}_0).
\end{equation}
For finding this probability, the distribution of the random variable $\mathcal{T}(Y^{[i]}_l) \in \mathbb{R}$ in \eqref{Threshold} should be derived. More precisely, the PFA for a predefined threshold can be written as:
\begin{equation}
{P}_{\rm FA}=1-{F}_{\vartheta^{[i]}_l}(\eta')
\end{equation}
where ${F}_{\vartheta^{[i]}_l}$ is the CDF of $\vartheta^{[i]}_l=\frac{1}{\sigma_z^2}Y^{[i]\dag}_{l}Y^{[i]}_l$, $\vartheta^{[i]}_l\sim \chi^2(T)$, and $\chi^2(T)$ is the central Chi-squared distribution with $T$ degrees of freedom. We can now rewrite $\mathcal{T}(Y^{[i]}_l)$ in \eqref{Threshold} as a function of the new variable $\vartheta^{[i]}_l$ as:
\begin{equation}
\mathcal{T}(Y^{[i]}_l) = \mathfrak{g}(\vartheta^{[i]}_l)=\frac{1}{\sigma_z^2\vartheta^{[i]}_l}\exp\left(\frac{\vartheta^{[i]}_l}{2T} \right).
\end{equation}
The function $\mathfrak{g}$ is a two-to-one function with a minimum at $\vartheta^{[i]}_l=2T\sigma_z^2$. An approximate inverse function of $\mathfrak{g}$ can be derived as \cite{Corless,Abramowitz1970}:
\begin{equation}
\mathfrak{g}^{-1}(\eta')=\left\{ \begin{array}{l}
-(2T){W}_{0}\left(\frac{-1}{(2T\sigma^2_z)\eta'}\right)\;\text{for}\;0<\vartheta^{[i]}_l<(2T)\\

-(2T){W}_{-1}\left(\frac{-1}{(2T\sigma^2_z)\eta'}\right)\;\;\text{for}\;\vartheta^{[i]}_l \geq (2T)
\end{array} \right. ,
\end{equation}
where ${W}_{0}\left(\cdot\right)$ and ${W}_{-1}\left(\cdot\right)$ are the lower and upper branches of Lambert W function, respectively, where the general form of ${W}(z)$ is defined by:
\begin{equation}F
z={W}(z)e^{{W}(z)}.
\end{equation}
The PFA can be finally written as:
\begin{align}
{P}_{\rm FA}=1&-\text{F}_{\vartheta^{[i]}_l}\left(\eta'\right)=1-\mathit{p}\left(\mathfrak{g}^{-1}_{\rm left}(\eta') \leq \vartheta^{[i]}_l <  \mathfrak{g}^{-1}_{\rm right}(\eta')\right) \\ \notag
=1  &- \mathcal{P}\left[-(2T){W}_{-1}\left(\frac{-1}{(2T\sigma^2_z)\eta'}\right),T\right]\\ \notag 
&+\mathcal{P}\left[-(2T){W}_{0}\left(\frac{-1}{(2T\sigma^2_z)\eta'}\right),T\right],
\end{align}
where $\mathcal{P}\left[\cdot,\cdot\right]$ stands for the regularized Gamma function which is defined as:
\begin{equation}
\mathcal{P}\left[x,k\right]=\frac{\gamma\left(x,k\right)}{\Gamma\left(x\right)},
\end{equation}
where $\Gamma\left(x\right)$ and $\gamma\left(x,k\right)$ are the Gamma and the lower incomplete Gamma functions, respectively. Note that each test statistic is evaluated to detect a certain data stream. Therefore, when there are more than one unused DoFs, the PFA analysis will be the same as each direction is considered independently.

\subsection{Advantages of Proposed Sensing Method}
The overall sensing scheme is summarized in Alg.~\ref{alg1}. As explained in Section IV, the DoF index search (second step) is performed only when the eigenvalue based method (first step) detects the presence of unused DoFs. There are two advantages in dividing the sensing process into two steps. First, the overall complexity of the sensing process is significantly reduced especially because there is no need to perform the complex DoF index search if no unused DoF is detected in the first step. The second reason for performing the eigenvalue based sensing in the first step is the ability of this method to sense the presence of spatial holes without the need for primary CSI. Therefore, the primary CSI is only acquired when the first sensing step flags the presence of spatial hole(s).

The complexity order of the eigenvalue based sensing is $\mathcal{O}(N^{[0]^2})$ multiplications for calculating the covariance matrix $\textbf{R}_{Y}$, while the eigenvalues can be calculated by at most $\mathcal{O}(N^{[0]^3})$ calculation in each sensing frame. On the other hand, the second phase of the proposed sensing algorithm requires several matrix operations. The number of multiplications for finding the sample vectors of all data streams is in the order of $\mathcal{O}(\mathcal{D}^4)$ where $\mathcal{D}=\sum{d^{[i]}}_{i=1}^{K}$. If we assume the number of antennas at the secondary receiver (fusion center) is equal to the total number of data streams, the order of required multiplications for the DoF index search is $\mathcal{O}({N^{[0]^4}})$. The test statistics $\mathcal{T}(Y^{[i]}_l)$ is also required to be calculated which is an exponential function. Moreover, for finding the threshold value in the proposed eigenvalue based method, the primary CSI is not required (only the covariance matrix ${\bf{\Sigma}}$ should be calculated).
%-------------------------------------------------------------------------------------------
\begin{algorithm}[tp]
  \caption{Proposed sensing algorithm}
  	\label{alg1}
  	\small
  	1: Calculate the sensing covariance matrix $\textbf{R}_{Y}$ using \eqref{eq:30}\\
  	2: Calculate the minimum eigenvalue of $\textbf{R}_{Y}$\\
    3: \textbf{while} $\lambda_{min}(\textbf{R}_{Y}) < \eta $\\
	4: ${\mkern 15mu}$\textbf{for} $\,\,\,i=1:K$\\
	5: ${\mkern 30mu}$\textbf{for} $\,\,\,l=1:d^{[i]}$\\
	6: ${\mkern 45mu}$Calculate the sensing vector $D_l^{[i]}$ using \eqref{eq:sensvec}\\
	7: ${\mkern 45mu}$Calculate the sample vector $Y_l^{[i]}$ \\
	8: ${\mkern 45mu}$Calculate the test statistics $\mathcal{T}(Y^{[i]}_l)$ using \eqref{Threshold}\\
	9: ${\mkern 45mu}$\textbf{if} $\,\,\,\mathcal{T}(Y^{[i]}_l)< \eta'$\\
	10: ${\mkern 60mu}$The $l^{th}$ data stream of the  $i^{th}$ user is unused\\
	11: ${\mkern 45mu}$\textbf{end if}\\
	12: ${\mkern 30mu}$\textbf{end for}\\
	13: ${\mkern 15mu}$\textbf{end for}\\
	14: ${\mkern 15mu}$Calculate the sensing covariance matrix $\textbf{R}_{Y}$ using \eqref{eq:30}\\
    15: ${\mkern 15mu}$Calculate the minimum eigenvalue of $\textbf{R}_{Y}$\\
	16: \textbf{end while}
	
      \normalsize
  \end{algorithm}
  %\vspace{-2cm}
%---------------------------------------------------------------------

\section{Numerical Results}
We consider a $3$-user IA system with two antennas in each node as the primary network and a secondary MIMO link as shown previously in Fig.~\ref{vector}. It is assumed that the design of IA precoders at the primary is done with the distributed numerical approach presented in \cite{Gomadam2011} with $20$ iterations. The secondary system applies the transmission scheme proposed in this paper. We assume that all the channel links are zero-mean unit-variance complex Gaussian distributed. We further assume that all the receive noise are zero mean complex Gaussian with unit variance. It is also considered that the same transmit power of $10$ dBW is used for both the primary and secondary systems in all scenarios. 
\subsection{Number of Secondary Antennas and Feasibility of CR}
\begin{figure}[!t]
\centering
%\vspa%ce{-0.25cm}
\includegraphics[width=3.3in]{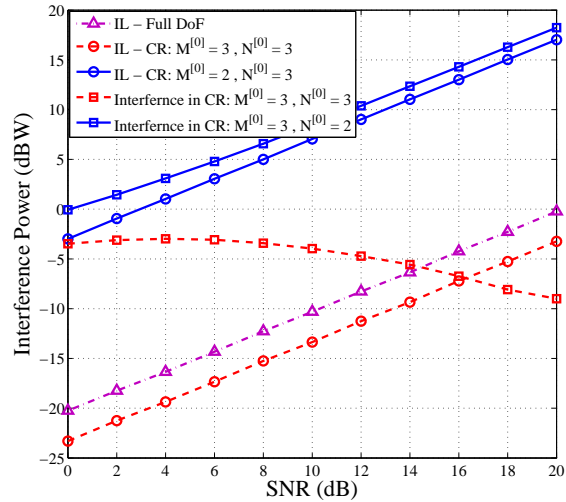}
\vspace{-0.5cm}
\caption{Interference power at the primary and secondary receivers for different scenarios when there is one inactive primary user.}
\vspace{-0.2cm}
\label{fig:inter1}
\end{figure}

Fig.~\ref{fig:inter1} shows the effect of the number of secondary antennas on the total interference leakage (IL) as a function of the SNR at the active primary receivers when there is only one inactive primary data stream. This figure also illustrates the interference from the primary transmitters to the secondary receiver for different number of secondary antennas. As seen, when the number of transmit antennas at the secondary is equal to or more than the total number of primary streams (in the figure, we have considered $N^{[0]}=3$), the total interference imposed to the primary network due to secondary transmission is less than the total interference at each primary receiver caused by other primary transmitters when all the primary pairs are active and there is no unused DoF. This means that applying our secondary transmission scheme does not increase the total interference power at the primary receivers. On the other hand, in the case of two transmit antennas at the secondary ($N^{[0]}=2$), the interference imposed to the primary is significantly larger, and the CR transmission is not feasible. This shows the significance of the inequality proved in \textit{Lemma 1} for the minimum required number of antennas at the secondary transmitter.

\begin{figure}[!t]
\centering
\includegraphics[width=3.3in]{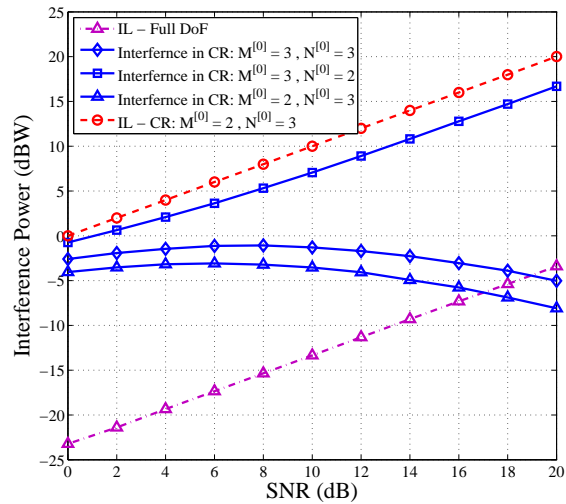}
\vspace{-0.5cm}
\caption{Interference power at the primary and secondary receivers for different scenarios when there are two inactive primary users.}
\vspace{-0.2cm}
\label{fig:inter2}
\end{figure}

\begin{figure}[t]
\centering
\includegraphics[width=3.3in]{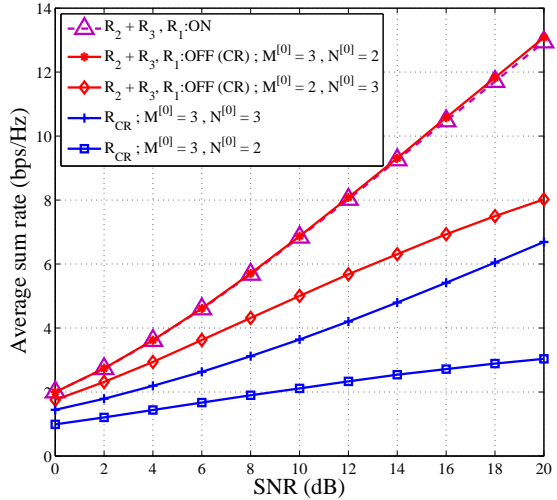}
\vspace{-0.5cm}
\caption{Average sum-rate of the primary system versus SNR before and after CR transmission when there is one inactive primary user. Figure also shows the average rate of the secondary for different antenna configurations.}
\vspace{-0.2cm}
\label{fig:sumrate1}
\end{figure}

The case of two inactive streams is illustrated in Fig.~\ref{fig:inter2}. We assume that the secondary network uses both of the spatial holes for its own transmission. This figure shows that the proposed method can be equally applied for opportunistic use of more than one spatial hole. In this case, since we consider two streams for the secondary network, the interference to the CR receiver increases when the number of antennas at CR transmitter is less than three. Since the interference leakage in the active primary receiver is almost zero for the three transmit antennas, we have not shown this curve in the figure.
%%%%%%%%%%%%%%%%%%%%%%%%%%%%%%%%%%%%%%%%%%%%%%%%%%%%%%%%%
%%%%%%%%%%%%%%%%%%%%%%%%%%%%%%%%%%%%%%%%%%%%%%%%%%%%%%%%%
%%%%%%%%%%%%%%%%%%%%%%%%%%%%%%%%%%%%%%%%%%%%%%%%%%%%%%%%%
%%%%%%%%%%%%%%%%%%%%%%%%%%%%%%%%%%%%%%%%%%%%%%%%%%%%%%%%%
On the other hand, the fundamental requirement for the secondary network is to be able to communicate at reasonable rate in the primary spectrum band. To see this, we have simulated the throughput of the secondary system in Fig.~\ref{fig:sumrate1}. This figure also shows the sum-rate performance of the primary system for different secondary antenna configurations. Based on this figure, when $M^{[0]}=3$, i.e. when the condition on the minimum number of antenna is satisfied, the secondary transmission, when primary transmitter one is off, does not affect the average sum-rate of pairs 2 and 3 of the primary network (i.e. $R_2+R_3$). However, this is not the case when $M^{[0]}=2$. In this case, i.e. when the number of transmit antennas at the secondary is less than the total DoFs of the primary, the sum-rate ($R_2+R_3$) of the primary degrades significantly due to the non-negligible secondary interference. This figure also shows that by increasing the number of receive antennas at the CR, its average rate increases significantly due to the secondary decoder.

\begin{figure}[t]
\centering
%\vspace{-0.25cm}
\includegraphics[width=3.3in]{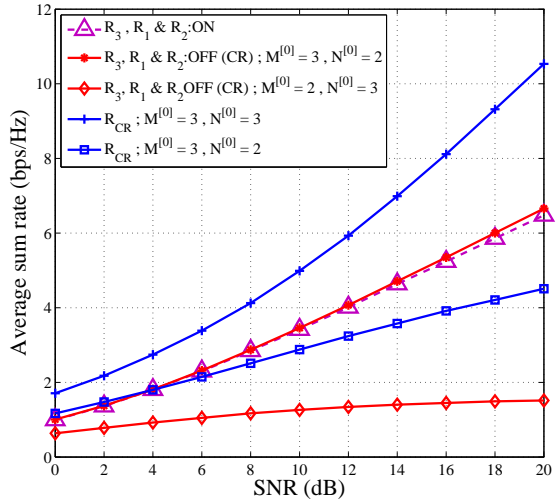}
\vspace{-0.5cm}
\caption{Average sum-rate of the primary system versus SNR before and after CR transmission when there are two inactive primary users. Figure also shows the average rate of the secondary for different antenna configurations.}
\vspace{-0.2cm}
\label{fig:sumrate2}
\end{figure}

In Fig.~\ref{fig:sumrate2}, we have simulated the sum rates for the case of two inactive primary streams, while the CR utilizes both available spatial holes for its own transmission. This figure shows that the secondary network does not affect the achievable rate of the primary system whenever the condition of \textit{Lemma 1} is met. On the other hand, the CR network is able to transmit two data streams and achieve a usable rate.

\begin{figure}[!t]
\centering
\includegraphics[width=3.3in]{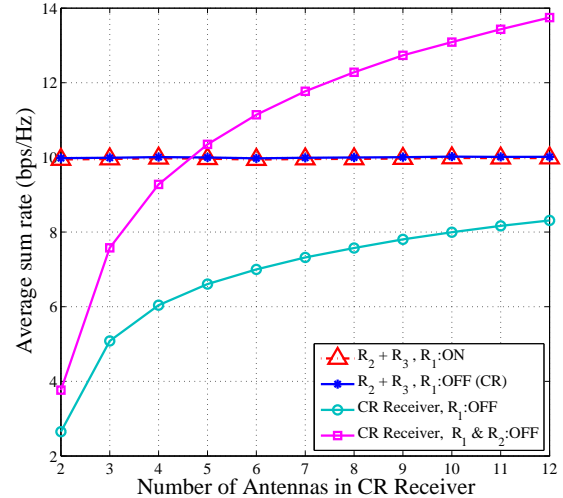}
\vspace{-0.5cm}
\caption{Average sum-rate of the primary system versus number of secondary receive antennas before and after CR transmission, $\text{SNR}=15\text{dB}$ and $M^{[0]}=3$.}
\vspace{-0.2cm}
\label{fig:CRantenna}
\end{figure}

Fig.~\ref{fig:CRantenna} illustrates the average rate of the secondary as well as the average sum-rate of the primary as a function of the number of secondary receive antennas. As shown while the number of secondary receive antennas does not have a significant effect on the sum-rate of the primary, the average secondary rate is always an increasing function of its number of receive antennas. This is in agreement with the  result of \textit{Lemma 3}. This figure also shows that the statement of the \textit{Lemma 3} is still valid for the case of more than one secondary data stream, and the self-interference does not change the behavior of average SINR with respect to number of antennas at the secondary. The figure also illustrates that after some points, increasing the number of secondary receive antennas has a small effect on its average rate. For example, according to Fig. ~\ref{fig:CRantenna}, increasing the number of antennas from 8 to 10 increases the average rate of the secondary by only 0.4 bps/Hz.
%%%%%%%%%%%%%%%%%%%%%%%%%%%%%%%%%%%%%%%%%%%%%%%%%%%%%%%%%
%%%%%%%%%%%%%%%%%%%%%%%%%%%%%%%%%%%%%%%%%%%%%%%%%%%%%%%%%
%%%%%%%%%%%%%%%%%%%%%%%%%%%%%%%%%%%%%%%%%%%%%%%%%%%%%%%%%
%%%%%%%%%%%%%%%%%%%%%%%%%%%%%%%%%%%%%%%%%%%%%%%%%%%%%%%%%

\subsection{Performance of Fast DoF Sensing Method}
\begin{figure}[!t]
\centering
%\vspace{-0.2cm}
\includegraphics[width=3.3in]{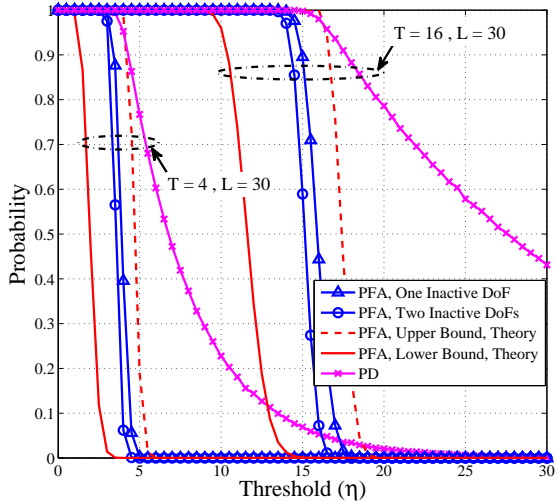}
\vspace{-0.5cm}
\caption{PFA, its lower and upper bounds, and the PD versus threshold for null space sensing.}
\vspace{-0.2cm}
\label{fig:fast}
\end{figure}

\begin{figure}[!t]
\centering
%\vspace{-0.2cm}
\includegraphics[width=3.3in]{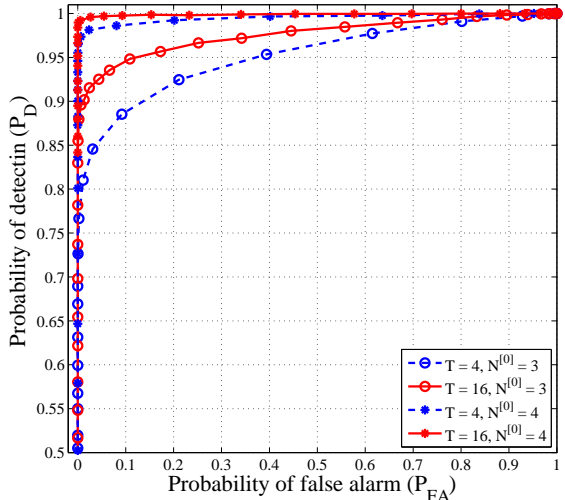}
\vspace{-0.5cm}
\caption{ROC curves of the null space sensing method for different smoothing factors and number of secondary receive antennas.}
\vspace{-0.2cm}
\label{fig:fast_roc}
\end{figure}

Fig.~\ref{fig:fast} shows the capability of the proposed eigenvalue based sensing scheme in detecting the presence of unused DoFs. We set the number of samples for producing the correlation matrix to $L=30$ to analyze the effect of number of samples on the performance of sensing method. The figure shows the PFA, and its theoretical upper and lower bounds as well as the correct detection probability as a function of threshold $\eta$ for two different values of smoothing factor $T$. As shown, at sufficiently large number of samples, the PFA suddenly drops while the PD is still high. Note that the theoretical upper bound of the PFA is fairly close to the Monte Carlo simulation results. This figure also shows that the PFA decreases when there are more than one spatial holes. To further demonstrate the capability of the proposed DoF sensing method in detecting the presence (or absence) of spatial hole(s), we evaluate its receiver operating characteristic (ROC) for different smoothing factors, $T$, and the number of secondary receive antennas, $N^{[0]}$. Fig. ~\ref{fig:fast_roc} shows the ROC of the proposed fast sensing method, and as seen, the ROC becomes sharper by increasing $T$ and $N^{[0]}$. The plot also shows that, especially for $N^{[0]}=4$, the probabilities of detection and false alarm are very close to one and zero, respectively. This means that the fast sensing method can reliably detect the presence and absence of spatial hole(s). Therefore, using it as a precursor step before performing DoF index search can significantly reduce the computational complexity at no or small hit in sensing performance.  

%%%%%%%%%%%%%%%%%%%%%%%%%%%%%%%%%%%%%%%%%%%%%%%%%%%%%%%%%
%%%%%%%%%%%%%%%%%%%%%%%%%%%%%%%%%%%%%%%%%%%%%%%%%%%%%%%%%
\subsection{Performance of DoF Index Search Method}
\begin{figure}[!t]
\centering
\includegraphics[width=3.3in]{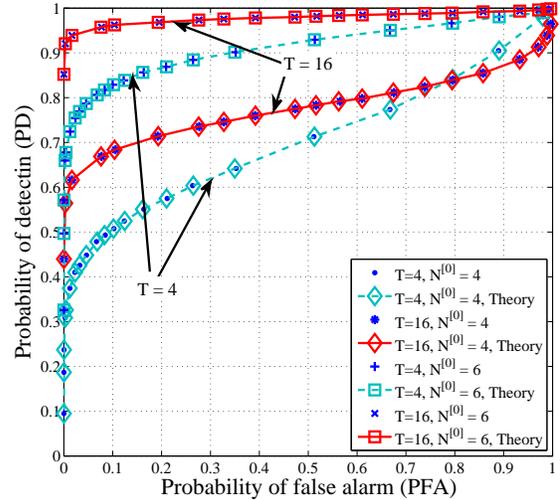}
\vspace{-0.5cm}
\caption{ROC curves of the DoF index search method for different smoothing factors and number of secondary receive antennas.}
\vspace{-0.2cm}
\label{fig:sens}
\end{figure}

The ROC performance results of the proposed searching method is shown in Fig.~\ref{fig:sens}. In this scheme, the number of sensing antennas is an important factor in finding the unused DoFs. It can be seen that, by increasing the number of secondary receive antennas, the PD for a specific PFA significantly increases, especially at a low PFA. As also shown in the figure, the experimental results matches perfectly with our theoretical analysis. The number of samples in the proposed GLRT detection is an important factor in the performance. As seen, the performance of the sensing scheme significantly increases with number of samples in each test. It can bee seen that for a PFA less than $0.3$, we can find a threshold that achieves a PD higher than $0.5$ to make the scheme practically feasible. 

%%%%%%%%%%%%%%%%%%%%%%%%%%%%%%%%%%%%%%%%%%%%%%%%%%%%%%%%%
%%%%%%%%%%%%%%%%%%%%%%%%%%%%%%%%%%%%%%%%%%%%%%%%%%%%%%%%%

%%%%%%%%%%%%%%%%%%%%%%%%%%%%%%%%%%%%%%%%%%%%%%%%%%%%%%%%%
%%%%%%%%%%%%%%%%%%%%%%%%%%%%%%%%%%%%%%%%%%%%%%%%%%%%%%%%%

\section{Conclusion}
In this paper, we addressed the feasibility of opportunistic DoF usage of a $K$-user interference alignment network. We showed that for the proper number of antennas at cognitive radio transmitter and receiver, the secondary is able to utilize the unused DoFs of the primary system to transmit its own information while causing zero or minima interference to the primary receivers. We then came up with proper secondary precoding and decoding matrices to make the secondary transmission possible. We then proposed a two-stage spatial DoF sensing approach. Using our scheme, the CR receiver is able to quickly detect the availability of unused primary DoFs or inactive streams. If a spatial hole is detected, in the second sensing stage, for finding the index set of inactive DoFs, the secondary system uses a spatial DoF index search method. With the help of simulations, we showed that the proposed opportunistic DoF usage scheme works well in practice and provides a significant throughput for the secondary system while causing no or minimal interference to the primary system.

%%%%%%%%%%%%%%%%%%%%%%%%%%%%%%%%%%%%%%%%%%%%%%%%%%%%%%%%%
%%%%%%%%%%%%%%%%%%%%%%%%%%%%%%%%%%%%%%%%%%%%%%%%%%%%%%%%%
\appendices

\section{Proof of Lemma 3}\label{app2}
To see the effect of the number of secondary receive antennas, $N^{[0]}$, on the average SINR, we show that by adding one additional antenna at the secondary receiver, the average SINR increases. We first define the new vector $\tilde{H}_l$ and matrix $\tilde{\textbf{B}}_l$ of the receiver with one additional antenna as:
\begin{align}
\tilde{H}_l=\left[ {\begin{array}{*{20}{c}}
{{H_l}}\\
{{h_l}}
\end{array}} \right], \tilde{\textbf{B}}_l=\left[ {\begin{array}{*{20}{c}}
{{\textbf{{B}}_l}}&{{B_1}}\\
{B_1^\dag }&{{b_2}}
\end{array}} \right],
\end{align}
where $\tilde{H}_l$ and $H_l$ are vectors of size $\left(N^{[0]}+1\right)\times 1$ and $N^{[0]}\times 1$, respectively, and matrices $\tilde{\textbf{B}}_l $ and $\textbf{{B}}_l$ are of size $\left(N^{[0]}+1\right)\times \left(N^{[0]}+1\right)$ and $N^{[0]}\times N^{[0]}$, respectively. Hence, the inverse of partitioned matrix $\tilde{\textbf{B}}_l$ can be written as:
\begin{equation}
\tilde{\textbf{B}}_l^{-1} = \left[ {\begin{array}{*{20}{c}}
{{{\hat{\textbf{B}}}^{ - 1}}}&{{{\hat B}_1}}\\
{\hat B_1^\dag }&{{{\hat b}_2}}
\end{array}} \right],
\end{equation}
where $\hat{\textbf{B}} = {\textbf{{B}}_l} - \frac{{{B_1}B_1^\dag }}{{{b_2}}}$, and $\hat{B}_1$ and $\hat{b}_2$ are well-defined vector and scalar, respectively \cite{Horn}. The average SINR at the secondary receiver in \eqref{SINR} can be written as:
\begin{align}\label{app1_1}
\E&\left[\tilde{\gamma}^{\rm CR max}_{l}\right]_{(N^{[0]}+1)}=\frac{p^{[0]}}{d^{[0]}}\E\left[\tilde{H}_l^{\dag}\tilde{\textbf{B}}_l^{-1}\tilde{H}_l\right] \\ \notag
&=\frac{p^{[0]}}{d^{[0]}}\left(\E\left[H_l^{\dag}\hat{\textbf{B}}^{-1}H_l\right]+2\Re\lbrace \E\left[h_l^*{\hat B}_1^{\dag}H_l\right]\rbrace+\E\left[|h_l|^2{\hat b}_2\right]\right).
\end{align}
Considering single stream transmission by the secondary, and based on the definition of interference matrix $\textbf{B}_l$ in (10), there is no common random variable in $\textbf{B}_l$ and $H_l$. Therefore, $h_l$, $H_l$ and ${\hat B}_1$ are independent, and the second term in \eqref{app1_1} is zero because $\E\left[h_l\right]=0$. Furthermore, the third term in \eqref{app1_1} is always positive. To prove the lemma, it suffices to compare the first term in this equation with the average SINR in the case of $N^{[0]}$ number of antennas.

For this purpose, we first prove that $\hat{\textbf{B}}$ is positive definite (PD); i.e. for all non-zero vectors $X$, $X^{\dag}\hat{\textbf{B}}X > 0$. Considering that matrix $\tilde{\textbf{B}}_l$ is PD, for any given non-zero vector $X^{H}=\left[ {\begin{array}{*{20}{c}} 
{{X_1^H}}&{{x_2^{*}}}
\end{array}} \right]$:
\begin{equation} \label{app1}
X^{\dag}\tilde{\textbf{B}}_lX = X_1^{\dag}\left(b_2 {\textbf{{B}}_l}-B_1B_1^{\dag}\right)X_1+||B_1^{\dag} X_1+b_2 x_2||^2.
\end{equation}
By choosing $x_2=-B_1^{\dag} X_1/b_2$, it follows that the first term in \eqref{app1}, i.e. $X_1^{\dag}\left(b_2 {\textbf{{B}}_l}-B_1 B_1^{\dag}\right)X_1$, is always positive. Therefore, the matrix $\left(b_2 {\textbf{{B}}_l}-B_1 B_1^{\dag}\right)$ is PD and hence, $\hat{\textbf{B}}$ is PD. Considering the fact that $\textbf{B}_l \succeq \hat{\textbf{B}}$  and since both matrices are PD, we can conclude that $\textbf{B}_l^{-1} \preceq \hat{\textbf{B}}^{-1}$ [32]. The first term in \eqref{app1_1} is always larger than the average SINR when the secondary receiver has $N^{[0]}$ antennas. So, $\E\left[\tilde{\gamma}^{\rm CR max}_{l}\right]_{(N^{[0]}+1)} \geq \E\left[{\gamma}^{\rm CR max}_{l}\right]_{(N^{[0]})}$, and the average SINR is an increasing function of the number of antennas at the secondary receiver.

\section{Proof of Lemma 5}\label{app3}
The first constraint of \eqref{optr} implies that the sensing vector $D_l^{[i]}$ must be orthogonal onto the subspace of matrix $\textbf{R}_{l}^{[i]}$ in $N^{[0]}$-dimensional space of the secondary receiver. The orthogonal projection of the desired vector $\textbf{H}^{[0i]}\textbf{V}^{[i]}_{\parallel l}$ onto subspace $\textbf{R}_{l}^{[i]}$ is \cite[5.13]{Projection}:
\begin{equation}\label{R}
C_l^{[i]}=\textbf{R}_l^{[i]}\left(\textbf{R}_l^{[i]\dag}\textbf{R}_l^{[i]}\right)^{-1}\textbf{R}_l^{[i]\dag}\textbf{H}^{[0i]}\textbf{V}^{[i]}_{\parallel l}.
\end{equation}
On the other hand, the objective function of \eqref{optr} is to find an orthogonal vector to $C_l^{[i]}$ with maximum inner product with the desired vector $\textbf{H}^{[0i]}\textbf{V}^{[i]}_{\parallel l}$. Therefore, the normalized vector $D_l^{[i]}$ should be in the direction of the orthogonal projector onto subspace $\textbf{R}_{l}^{[i]}$ and can be written as:
\begin{equation}
D_l^{[i]}=\frac{1}{\lambda}\left(\textbf{H}^{[0i]}\textbf{V}^{[i]}_{\parallel l}-C_l^{[i]}\right),
\end{equation}
where normalization parameter $\lambda$ is:
\begin{equation}
\lambda^2=\left(\textbf{V}^{[i]\dag}_{\parallel l}\textbf{H}^{[0i]\dag}-C_l^{[i]\dag}\right)\left(\textbf{H}^{[0i]}\textbf{V}^{[i]}_{\parallel l}-C_l^{[i]}\right).
\end{equation} 

\bibliographystyle{IEEEtran}
\bibliography{IEEEabrv,references}
\vspace{5cm}
\begin{biography}
[{\includegraphics[width=1in,height=1.25in,clip,keepaspectratio]{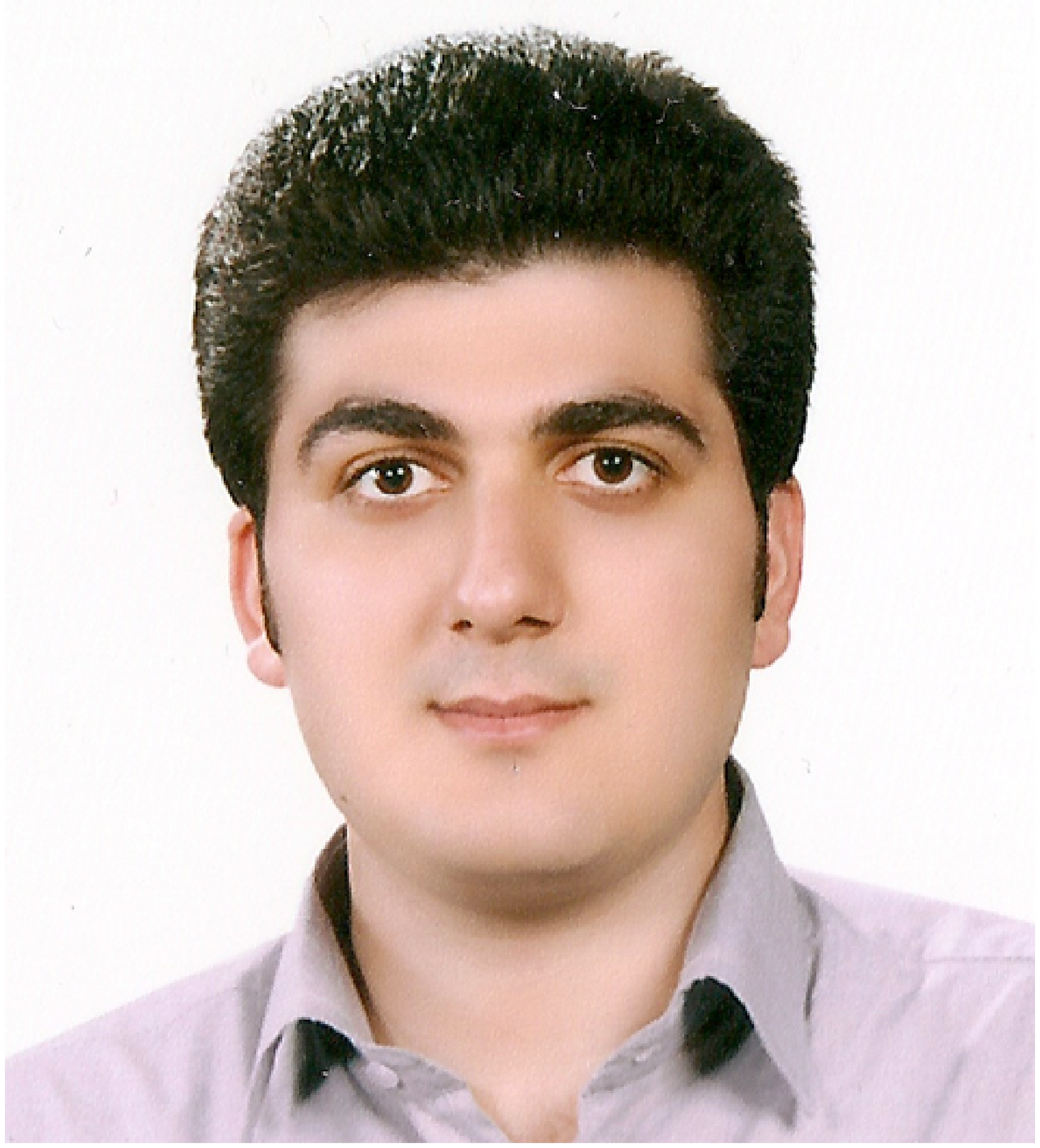}}]{Ardalan Alizadeh}
is currently a Ph.D. student at the Department of Electrical and Computer Engineering, the University of Akron, OH, USA. He received his B.Sc. and M.Sc. degrees from Amirkabir University of Technology and Shahid Beheshti University, Tehran, Iran in 2008 and 2011, respectively. He is also a Research Assistant in wireless communications laboratory (WCL) of the Department of Electrical and Computer Engineering, the University of Akron. His current area of research includes cognitive radio networks, interference alignment in MIMO interference networks, cooperative communications, spatial modulation and implementation of wireless communication systems by software defined radio (SDR). He regularly serves as the reviewer of IEEE conferences and journals including IEEE GLOBECOM, International Conference on Communications (ICC) and IEEE Transactions on Communications. 
\end{biography}
\vspace{-2cm}
\begin{biography}
[{\includegraphics[width=1.08in,height=1.35in,clip,keepaspectratio]{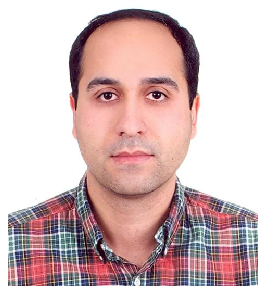}}]{Hamid Reza Bahrami}
is currently an Assistant Professor at the Department of Electrical and Computer Engineering at The University of Akron, OH, USA. His main area of research includes wireless communication, information theory, and applications of signal processing in communication. He received his Ph.D. degree in electrical engineering from McGill University, Montreal, Canada in 2008. From 2007 to 2009 and prior to joining the University of Akron, he was a scientist at Wavesat Inc. Montreal, Canada. Dr. Bahrami has B.Sc. and M.Sc. degrees both in Electrical Engineering from Sharif University of Technology and University of Tehran, Iran, respectively. He has served as the Editor for Transactions on Emerging Telecommunications Technologies, as the Guest Editor for The Scientific World Journal and as the Technical Program Committee member of numerous IEEE conferences including IEEE GLOBECOM and International Conference on Communications (ICC). He is currently a member of the IEEE, the IEEE Communications Society, and the IEEE Vehicular Technology Society.
\end{biography}
\vspace{-2cm}
\begin{biography}
[{\includegraphics[width=1in,height=1.25in,clip,keepaspectratio]{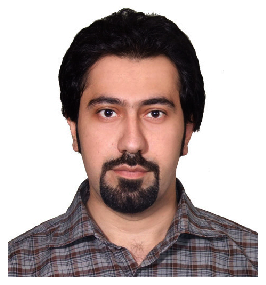}}]{Mehdi Maleki}
received the B.Sc. and M.Sc. degrees in electrical engineering from Amirkabir University of Technology (Tehran Polytechnic), Tehran, Iran, in 2006 and 2009, respectively. He is currently working toward the Ph.D. degree with The University of Akron, OH, USA. He is also a Research Assistant in wireless communications laboratory (WCL) of the Department of Electrical and Computer Engineering, The University of Akron. He regularly serves as reviewer for IEEE transactions/journals and major conferences. His research interests include digital communications, digital
signal processing, multi-user communications, multiple-input multiple-output wireless systems, cooperative communications and cognitive radio networks.
\end{biography}
\vspace{-2cm}
\begin{IEEEbiography}[{\includegraphics[width=1in,height=1.25in, clip, keepaspectratio]{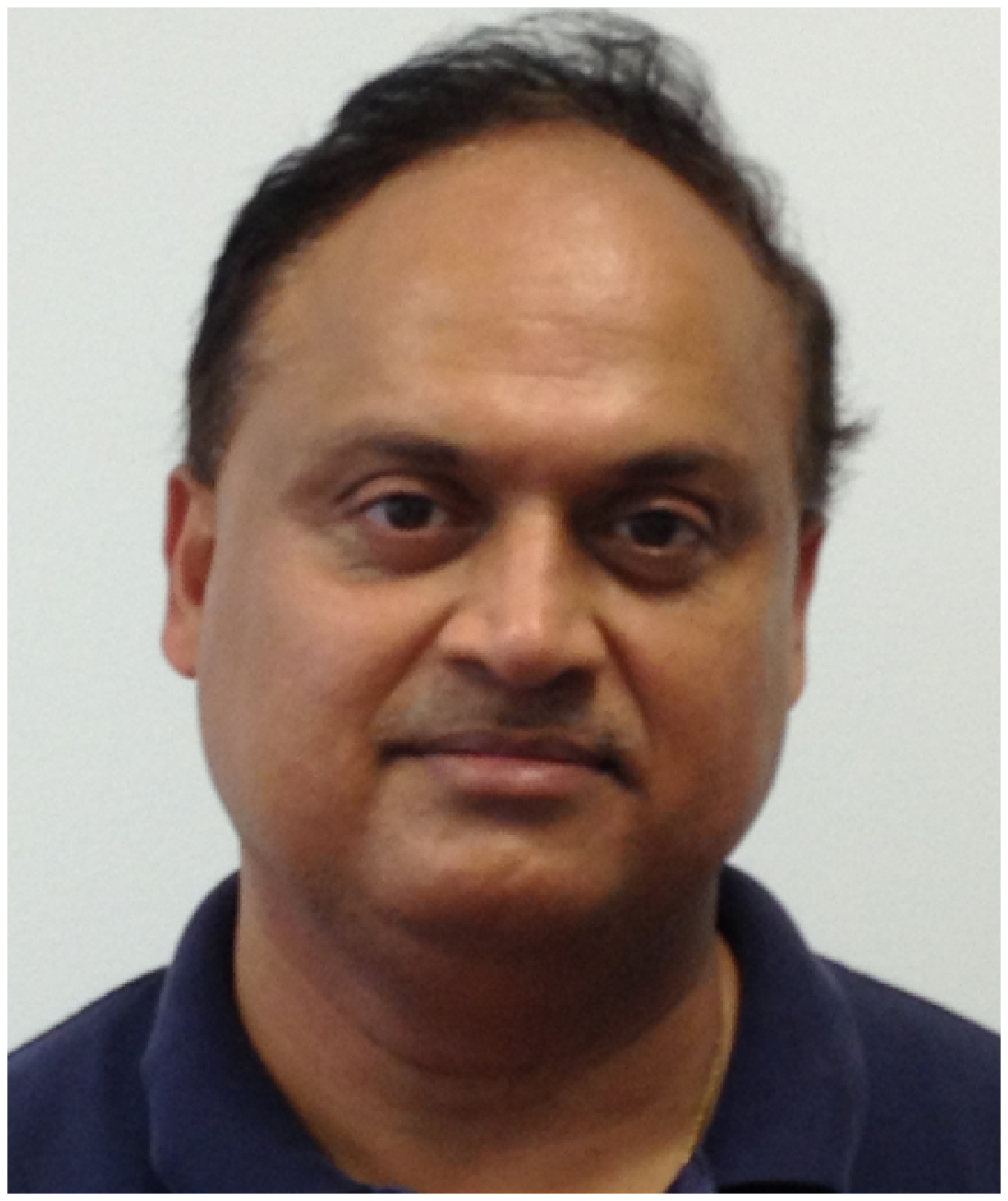}}]
{Shivakumar Sastry} is an Associate Professor with the Department of Electrical and Computer Engineering, The University of Akron. He received his Ph.D. degree in Computer Engineering and Science from Case Western Reserve University and holds Masters Degrees in Computer Science from University of Central Florida and in Electrical Engineering from the Indian Institute of Science.  His research interests are in Networked Embedded Systems, Real-time systems, and Graph algorithms. Prior to joining Akron, he was a Senior Research Scientist with Rockwell Automation.
\end{IEEEbiography}

\end{document}